\documentclass[twocolumn]{aastex631} 
\PassOptionsToPackage{svgnames}{xcolor}
\usepackage{graphicx}	% Including figure files
\usepackage{amsmath}	% Advanced maths commands
\usepackage{amssymb}	% Extra maths symbols
\usepackage{bm}		% Bold maths symbols, including upright Greek
\usepackage{float}
\usepackage{natbib}
\usepackage{scalerel}
\usepackage{comment}
\usepackage{hyperref}
\usepackage{changepage}
\usepackage{color,soul}
\usepackage{txfonts}
\usepackage{tikz}
\newcommand{\agel}{{\tt AGEL}}

\newcommand{\lcdm}{$\Lambda$CDM}
\newcommand{\wcdm}{$w$CDM}
\newcommand{\om}{$\Omega_{\rm m}$}
\newcommand{\ode}{$\Omega_{\rm \Lambda}$}
\newcommand{\ok}{$\Omega_{k}$}
\newcommand{\w}{$w$}
\newcommand{\wo}{$w$}

\newcommand{\zdefl}{$z_{\rm defl}$}

\newcommand{\zone}{$z_{\scalebox{0.5}{S1}}$}
\newcommand{\ztwo}{$z_{\scalebox{0.5}{S2}}$}

\begin{document}
\shorttitle{Cosmology with DSPLs}
\shortauthors{Sahu et al.}
\defcitealias{Tran:Harshan:2022}{AGEL-DR1}
\submitjournal{The Astrophysical Journal}
%\received{}
%\revised{}
%\accepted{}

%\graphicspath{{./}{Figures/}}
\maxdeadcycles=100000
%%%%%%%%%%%%%%%%%%% TITLE PAGE %%%%%%%%%%%%%%%%%%%

\title{Cosmography with the Double Source Plane Strong Gravitational Lens \agel150745+052256}

\correspondingauthor{Nandini Sahu}
\email{n.sahu@unsw.edu.au}

\author[0000-0003-0234-6585]{Nandini Sahu}
\affil{Center for Astrophysics, Harvard \& Smithsonian, Cambridge, MA 02138, USA}
\affil{University of New South Wales, NSW 2052, Australia}
\affil{The ARC Centre of Excellence for All Sky Astrophysics in 3 Dimensions (ASTRO 3D), Australia}

\author[0000-0002-5558-888X]{Anowar J.~Shajib}\thanks{NHFP Einstein Fellow}
\affil{Department of Astronomy \& Astrophysics, University of Chicago, Chicago, IL 60637, USA}
\affil{Kavli Institute for Cosmological Physics, University of Chicago, Chicago, IL 60637, USA}
\affil{Center for Astronomy, Space Science and Astrophysics, Independent University, Bangladesh, Dhaka 1229, Bangladesh}

\author[0000-0001-9208-2143]{Kim-Vy Tran}
\affil{Center for Astrophysics, Harvard \& Smithsonian, Cambridge, MA 02138, USA}
\affil{The ARC Centre of Excellence for All Sky Astrophysics in 3 Dimensions (ASTRO 3D), Australia}

\author{Hannah Skobe}
\affil{McWilliams Center for Cosmology and Astrophysics, Physics Department, Carnegie Mellon University, Pittsburgh, PA 15213, USA}
\affil{Department of Astronomy \& Astrophysics, University of Chicago, Chicago, IL 60637, USA}

\author{Sunny Rhoades}
\affil{Department of Physics and Astronomy, University of California, Davis, 1 Shields Avenue, Davis, CA 95616, USA}

\author[0000-0001-5860-3419]{Tucker Jones}
\affil{Department of Physics and Astronomy, University of California, Davis, 1 Shields Avenue, Davis, CA 95616, USA}

\author{Karl Glazebrook}
\affil{Centre for Astrophysics and Supercomputing, Swinburne University of Technology, PO Box 218, Hawthorn, VIC 3122, Australia}
\affil{The ARC Centre of Excellence for All Sky Astrophysics in 3 Dimensions (ASTRO 3D), Australia}

\author{Thomas E. Collett}
\affil{Institute of Cosmology and Gravitation, University of Portsmouth, Burnaby Rd, Portsmouth, PO1 3FX, United Kingdom}

\author{Sherry H.~Suyu}
\affil{Technical University of Munich, TUM School of Natural Sciences, Physics Department,  James-Franck-Stra{\ss}e 1, 85748 Garching, Germany}
\affil{Max-Planck-Institut f{\"u}r Astrophysik, Karl-Schwarzschild Stra{\ss}e 1, 85748 Garching, Germany}

\author{Tania M. Barone}
\affil{University of New South Wales, Sydney, NSW, Australia}
\affil{Centre for Astrophysics and Supercomputing, Swinburne University of Technology, PO Box 218, Hawthorn, VIC 3122, Australia}
\affil{The ARC Centre of Excellence for All Sky Astrophysics in 3 Dimensions (ASTRO 3D), Australia}

\author[0000-0002-2645-679X]{Keerthi Vasan G.C.}
\affil{The Observatories of the Carnegie Institution for Science, 813 Santa Barbara Street, Pasadena, CA 91101, USA}

\author{Duncan J. Bowden}
\affil{Center for Astrophysics, Harvard \& Smithsonian, Cambridge, MA 02138, USA}
\affil{School of Physics \& Astronomy, University of Southampton, Southampton SO17 1BJ, UK}

\author[0009-0003-3198-7151]{Daniel Ballard}
\affil{Sydney Institute for Astronomy, School of Physics A28, University of Sydney, NSW 2006, Australia}

\author[0000-0003-1362-9302]{Glenn G. Kacprzak}
\affil{Centre for Astrophysics and Supercomputing, Swinburne University of Technology, PO Box 218, Hawthorn, VIC 3122, Australia}
\affil{The ARC Centre of Excellence for All Sky Astrophysics in 3 Dimensions (ASTRO 3D), Australia} 

\author[0000-0002-1576-2505]{Sarah M. Sweet}
\affil{School of Mathematics and Physics, University of Queensland, Brisbane, QLD 4072, Australia}
\affil{The ARC Centre of Excellence for All Sky Astrophysics in 3 Dimensions (ASTRO 3D), Australia}

\author[0000-0003-3081-9319]{Geraint F. Lewis}
\affil{Sydney Institute for Astronomy, School of Physics A28, The University of Sydney, NSW 2006, Australia}

\author[0000-0003-2804-0648 ]{Themiya Nanayakkara}
\affil{Centre for Astrophysics and Supercomputing, Swinburne University of Technology, PO Box 218, Hawthorn, VIC 3122, Australia}

\nocollaboration{17}


\begin{abstract}

Strong gravitational lenses with two background sources at widely separated redshifts are a powerful and independent probe of cosmological parameters.
We can use these systems, known as Double-Source-Plane Lenses (DSPLs), to measure the ratio ($\beta$) of angular-diameter distances of the sources, which is sensitive to the matter density (\om) and the equation-of-state parameter for dark-energy (\wo). However, DSPLs are rare and require high-resolution imaging and spectroscopy for detection, lens modeling, and measuring $\beta$. 
Here we report only the second DSPL ever used to measure cosmological parameters. We model the DSPL \agel150745+052256 from the ASTRO~3D Galaxy Evolution with Lenses (\agel) survey using HST/WFC3 imaging and Keck/KCWI spectroscopy. 
The spectroscopic redshifts for the deflector and two sources in \agel1507 are \zdefl$=0.594$, \zone$=2.163$, and \ztwo$=2.591$. We measure a stellar velocity dispersion of $\sigma_{\rm obs}=109\pm27$~km s$^{-1}$ for the nearer source (S1). Using $\sigma_{\rm obs}$ for the main deflector (from literature) and S1, we test the robustness of our DSPL model. We measure $\beta=0.953^{+0.008}_{-0.010}$ for \agel1507 and infer \om$=0.33^{+0.38}_{-0.23}$ for \lcdm\ cosmology. Combining \agel1507 with the published model of the Jackpot lens improves the precision on \om\ (\lcdm) and \wo\ (\wcdm) by $\sim10$\%. The inclusion of DSPLs significantly improves the constraints when combined with \textit{Planck}’s cosmic microwave background observations, enhancing precision on \wo\ by 30\%. This paper demonstrates the constraining power of DSPLs and their complementarity to other standard cosmological probes. Tighter future constraints from larger DSPL samples discovered from ongoing and forthcoming large-area sky surveys would provide insights into the nature of dark energy.

\end{abstract}

\section{Introduction}
\label{Sec:Introduction}

The $\Lambda$ Cold Dark Matter 
(\lcdm) model is currently the most widely accepted model of the Universe.
In the \lcdm \ model, baryonic and dark matter account for about $30\%$ (\om$\approx 0.3$) of present day energy density of the Universe, the remaining $70\%$ is accounted by dark energy (\ode$\approx 0.7$) that does not evolve with time ($w=-1$), and the geometry of the Universe is flat (\ok$= 0$).
Observations of cosmic microwave background \citep[CMB,][]{Planck:2020}, baryonic acoustic oscillations \citep[BAO,][]{Alam:Aubert:2021}, and Type Ia supernovae \citep[SNe,][]{Scolnic:2018} suggest that the formation of large scales structures ($\gtrsim \rm Mpc$) in the Universe are very well described by the \lcdm \ model.

%At smaller scales  ($< \rm 1 \ Mpc$), \lcdm\ faces many challenges where observations do not match the predictions by the \lcdm \ model  \citep[see reviews in][]{DelPopolo:2017, Bullock:BoylanKolchin:2017, Salucci:review:2019}. Some famous open questions  are, the lack of observed massive dark matter dominated sub halos \citep[too big to fail problem,][]{Boylan-Kolchin:Bullock:2011}, the sparseness of low-mass satellite galaxies, i.e., dark matter sub-halos in the Local Group \citep[the missing satellites problem,][]{Klypin:Kravtsov:1999, Nashimoto:2022}, and the observations of flat (less dense) cores of dwarf and low surface brightness galaxies against the predicted high density steep (cuspy) cores predicted by the \lcdm \ model \citep[the core/cusp problem,][]{Flores:Primack:1994, Moore:1994}.

At smaller scales ($< \rm 1 \ Mpc$), the \lcdm\ model faces several challenges where observations do not align with its predictions—such as the too big to fail  problem, the missing satellites problem, and the core/cusp problem \citep[see reviews in][]{DelPopolo:2017, Bullock:BoylanKolchin:2017, Salucci:review:2019}. While recent observations from wide-field surveys \citep{Drlica-Wagner:Bechtol:2015, Homma:Chiba:2024}, combined with sample completeness corrections \citep{Kim:Peter:2018} and improved simulations \citep{Brooks:Kuhlen:2013, Fielder:Mao:2019}, appear to have addressed the missing satellites problem, many discrepancies between observations and \lcdm\ predictions remain unresolved \citep[see][]{Perivolaropoulos:Skara:2022}.

Discrepancies observed between various distance probes for the expansion rate, $H_0$, of the Universe \citep{Verde:Treu:Riess:2019, Planck:2020},  
inconsistency of the dark-energy equation-of-state parameter, \w, observed from the combined BAO, CMB, and SNe \citep{DESI:DR2:BAO:2025} data with that of \lcdm\ model,
and alternative dark energy models \citep{Motta:2021} favored by current observational data \citep{Shajib:Frieman:2025, Giare:Mahassen:2025}, suggest that the standard \lcdm \ model of the Universe needs further independent testing and possibly modifications.

Gravitational lenses form independent cosmological probes due to the sensitivity of observed lensing morphology to the distances between the lens, background source, and the observer \citep{Blandford:Narayan:1992}. 
One famous example is the use of gravitational lenses
as the geometric probe of the expansion rate of the Universe  \citep[$H_{0}$,][]{Refsdal:1964}. This is done using the time delay between multiple lensed images of photometrically variable background sources such as quasars or supernovae  \citep{Suyu:Marshall:2010,Shajib:Birrer:DSPL:2020, Birrer:Millon:2024,Suyu:2024}. 
Another important application of gravitational lenses, that this paper will focus on, is constraining cosmological parameters, such as the matter density (\om), dark energy density (\ode), curvature of the Universe (\ok), and the dark-energy equation-of-state parameter ($w$), independent of the Hubble's constant.

Galaxy-scale lenses, which have simple deflector mass distribution and fewer perturbations than a galaxy group/cluster, can be powerful probes of cosmology \citep{Li:Collett:2024}. 
Studies suggest that cosmological models can be tested simply by measuring the Einstein radius and the enclosed total mass \citep{Biesiada:Piorkowska:2010}. However, using lensing-only data, the inferred deflector mass is degenerate with the deflector mass profile; thus, additional information about the mass model, such as deflector stellar kinematics, is required for robust cosmological inference.

Galaxy-scale double-source-plane lenses (DSPLs) with two background sources at widely separated redshifts are expected to be particularly sensitive to cosmology \citep{Collett:Auger:2012, Sharma:Linder:2022}. This is because, in a DSPL, the ratio of deflection angles for the two sources is equal to a distance ratio, $\beta$, involving the angular-diameter distances of the two background sources from the deflector and the observer. As the angular-diameter distance depends on redshifts and cosmology, an independent measurement of $\beta$ can constrain cosmological parameters when the redshifts are independently known.

Importantly, in DSPLs, the farther source provides additional constraints for a robust measurement of the deflector’s mass distribution and, therefore, $\beta$.
By extension, this also applies to lenses with more than two source planes. 
However, galaxy-scale lenses with two or more sources at different redshifts are extremely rare \citep[$<1\%$ of total lens population,][]{Ferrami:Wyithe:2024}. 
To date only $O(10)$ such lenses have been discovered in various surveys \citep{ Gavazzi:Treu:2008, Tu:Gavazzi:2009, Tanaka:Wong:2016, Schuldt:2019, Shajib:Birrer:DSPL:2020, Dux:Millon:2025, 
AGEL:DR2:2025, Euclid:Walmsley:2025}. 

Using the Jackpot lens, SDSSJ0946+1006 (hereafter J0946), \citet{Collett:Auger:2014} showed for the first time that a DSPL with known deflector and source redshifts can infer cosmological parameters such as \om\ and $w$, independent of $H_{0}$ \citep[see also][]{Gavazzi:Treu:2008}. Importantly, they found that cosmology constraints from DSPL J0946 are orthogonal to those from the CMB observations and improve the combined constraints by $30\%$ compared to CMB constraints alone. 

Cluster lenses with multiple lensed sources can  also be used for this purpose \citep{Link:Pierce:1998, Golse:Kneib:2002, Caminha:Suyu:2022}.
However, cluster-scale lenses have a complex mass profile, involving multiple components in the deflector plane. Therefore, the cluster mass models have a higher uncertainty than galaxy-scale lenses, which might result in a higher uncertainty about the inferred cosmological parameters.

In this paper, we present lens modeling and cosmological constraints inferred from a new galaxy-scale DSPL, \agel150745+052256 (hereafter \agel1507),  discovered in the ASTRO 3D Galaxy Evolution with Lenses survey\footnote{ AGEL survey website \href{https://sites.google.com/view/agelsurvey/}{https://sites.google.com/view/agelsurvey/}} \citep[\agel,][]{Tran:Harshan:2022, AGEL:DR2:2025}.
\agel1507 is the second ever DSPL used for cosmological inference.
We present the velocity dispersion measurement of the nearer source using integral field spectra and test our lens model by comparing 
model-predicted and observed velocity dispersions for the deflector and the nearer source.

This paper aims to present the second case study of cosmography with DSPLs, demonstrating the application of DSPLs to infer \om\ and \wo\ independently, and highlighting their complementarity with standard cosmological probes such as the CMB, BAO, and SNe Ia. Furthermore, we combine the constraints from the two DSPLs, \agel1507 and J0946, and analyze the significance of the combined inference, showing the importance of discovering DSPLs in next-generation wide-area sky surveys such as the \textit{Euclid} Wide Survey \citep{Euclid:Walmsley:2025} and the Rubin Observatory’s Legacy Survey of Space and Time \citep[LSST,][]{Shajib:Rubin-forecast:2024}.

This paper is arranged as follows. 
The multi-source plane gravitational lensing formalism is described in \autoref{Sec:compound lens theory}. 
The imaging and spectroscopy used in this work are presented in \autoref{Sec:data}. 
The lens modeling of \agel1507 is detailed in \autoref{Sec:modeling}, and the modeling results are presented in \autoref{Sec:modeling results}. \autoref{Sec:Comological inferences} presents the cosmological constraints based on our model of the DSPL \agel1507, as well as the combined constraints with J0946 and other cosmological probes. 
Finally, \autoref{Sec:conclusions} summarizes our findings and  the future scope of this work.

\section{Compound Gravitational Lensing}
\label{Sec:compound lens theory}

Gravitational lensing is the deflection of light from a background source by a foreground mass concentration along the same line of sight. 
Strong lensing, where multiple distorted and magnified images of a background source are formed, is caused by massive objects such as galaxies (with total mass $M=O(10^{13} M_\odot)$) or clusters ($M=O(10^{15} M_\odot)$).
The observed lensing morphology, i.e., the separation between lensed images, depends on the mass distribution of the deflector and the angular diameter distances between the deflector, source, and the observer \citep[see][for a review]{Saha:Sluse:2024}.

\begin{figure*}
    \centering
    \includegraphics[clip=true,trim= 0mm 0mm 0mm 0mm, width=  0.8\textwidth]
   {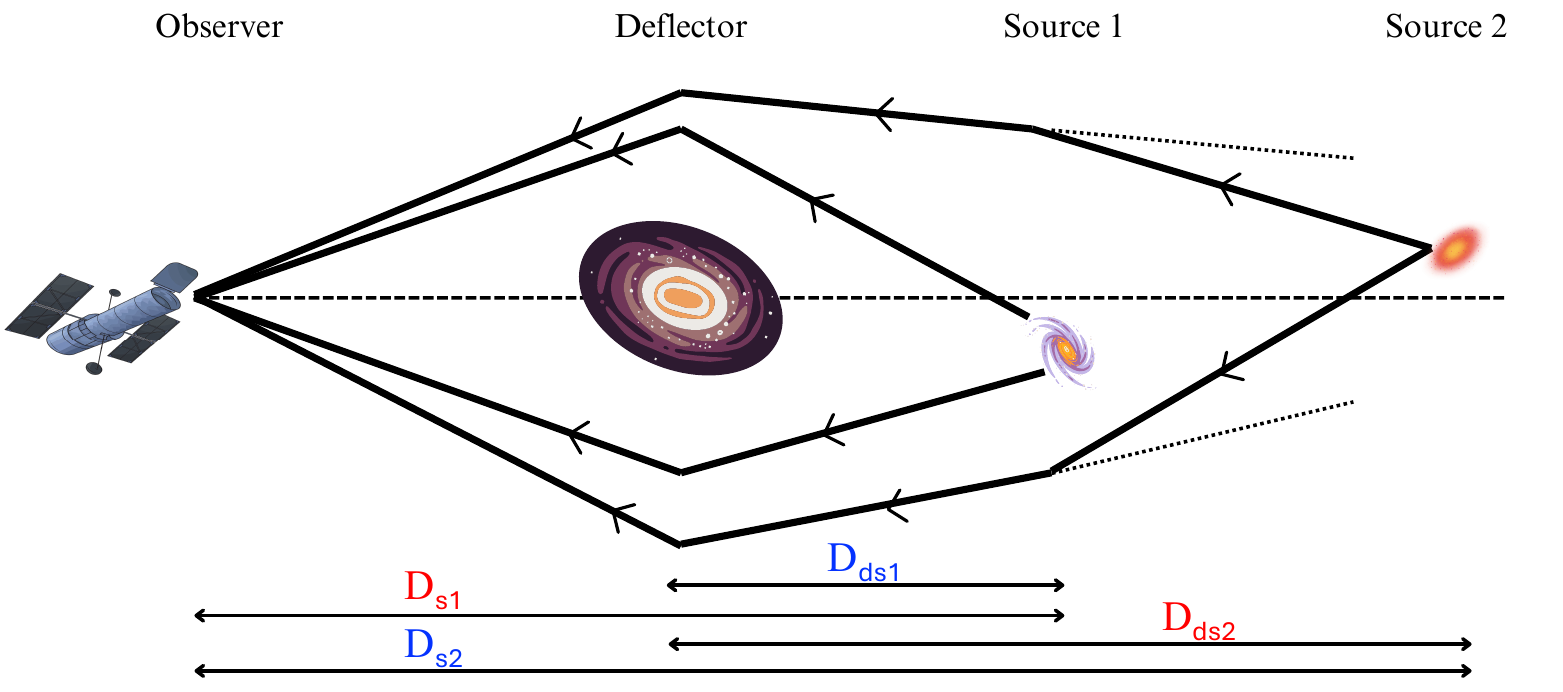}
    \caption{Schematic of a double-source-plane lens. Double-headed arrows at the bottom represent the angular diameter distances of the deflector and the sources. Blue and red colors of the text mark distances that are in the numerator and denominator, respectively, of the cosmological scaling factor, $\beta_{12}$ (\autoref{eq:beta_12}). 
    }
    \label{fig:ray_diag}
\end{figure*}

For a single-source-plane lens, the lens equation  that links the deflector plane coordinates, $\bm{\theta}$, with that of background source plane, $\bm{\theta_{\rm s}}$, via the scaled deflection angle $\bm{\alpha}(\bm{\theta}) = \frac{D_{\rm ds}}{D_{\rm s}}  \bm{\hat{\alpha}} (\bm{\theta})$, is expressed as 
\begin{equation}
    \label{eq:lens eq single plane}
  \bm{\theta_{\rm s}} = \bm{\theta} -\frac{D_{\rm ds}}{D_{\rm s}}  \bm{\hat{\alpha}} (\bm{\theta})  
\end{equation}
Here $ \bm{\hat{\alpha}} (\bm{\theta})$ represents the physical deflection angle as the light coming from the source crosses the plane of the deflector $\bm{\theta}$. $D_{\rm ds} $ and $D_{\rm s} $ are angular diameter distances between the deflector and source and between observer and source, respectively.
The scaled deflection angle is related to the gradient of the deflector potential as $\bm{\alpha}(\bm{\theta})=\nabla \psi(\bm{\theta})$.

\subsection{Multi-plane Gravitational Lens System}
\label{subsec: multi plane lens formalism}

In a multi-plane lens system, the light coming from the farthest source is deflected by all the objects along the line of sight, as depicted in \autoref{fig:ray_diag} as a ray diagram for DSPLs. 
For such a scenario, \citet{Schneider:Ehlers:Falco:1992} modified the lens equation to a recursive multi-plane lens equation accounting for the compounded lensing.
The multi-plane lens equation, with the redshift plane nearest to the observer at index $i=1$, can be expressed as
\begin{equation}
    \label{eq:lens eq multi-plane}
    \bm{\theta_{j}} = \bm{\theta_{1}} - \sum_{i=1}^{j-1} \beta_{ij} \bm{\alpha_{i}^{\prime}}(\bm{\theta_{i}}).
\end{equation}
It relates the angular position on $j^{\rm th}$ plane, $\bm{\theta_{j}}$, to the angular position on the nearest plane ($\bm{\theta_{1}}$) and compounded scaled deflection ($\sum_{i=1}^{j-1} \beta_{ij} \bm{\alpha_{i}^{\prime}}(\bm{\theta_{i}})$) by all the planes prior to the $ j^{\rm th}$ plane.
Here, $\beta_{ij}$ is the cosmological scaling factor defined as
\begin{equation}
    \beta_{ij} = \frac{D_{ij} D_{n}}{D_{j} D_{in}}
\end{equation}
and $\bm{\alpha_{i}^{\prime}}(\bm{\theta_{i}})$ represents the physical deflection angle rescaled for the final source represented with index $n$
\begin{equation}
    \bm{\alpha_{i}^{\prime}}(\bm{\theta_{i}})=\frac{D_{in}}{D_{n}} \bm{\hat{\alpha}_{i}} (\bm{\theta})
\end{equation}

Following \autoref{eq:lens eq multi-plane}, for  
a DSPL with one deflector plane and two source planes,  the lens equation pertaining to the planes of the two source galaxies can be expressed as
\begin{equation}
    \bm{\theta_{2}} = \bm{\theta_{1}} -\beta_{12}  \bm{\alpha^{\prime}_{1}}(\bm{\theta_{1}}), 
\end{equation}
and 
\begin{equation}
    \bm{\theta_{3}} = \bm{\theta_{1}} -  \bm{\alpha^{\prime}_{1}}(\bm{\theta_{1}})
    - \bm{\alpha^{\prime}_{2}}(\bm{\theta_{1}} -\beta_{12}  \bm{\alpha^{\prime}_{1}}(\bm{\theta_{1}}))
\end{equation}
with indices 1, 2, and 3 for the deflector plane, the nearer source, and the farthest source, respectively.

\subsection{Cosmological scaling factor $\beta$}
\label{subsec: Beta}
For a DSPL, the cosmological scaling factor $\beta$ is given by 
\begin{equation}
    \beta_{12}= \frac{D_{\rm d s1} D_{\rm s2}}{D_{\rm s1} D_{\rm d s2}} 
    \label{eq:beta_12}
\end{equation}
where, $D_{\rm s1}$, $D_{\rm d s1}$,  $D_{\rm s2}$, and $D_{\rm d s2}$
are angular diameter distances between the observer and source 1 (nearer source), deflector and source 1, observer and source 2 (farther source), and deflector and source 2, respectively.
Each distance measure is a function of redshift and cosmological parameters such that
\begin{equation}
    D_{ij}=\frac{c}{H_{0} (1+z_j)} \int_{z_i}^{z_{j}} \frac{dz}{E(z)} 
\end{equation}
for a flat Universe, i.e., $\Omega_k =0$.
Here, $E(z) \equiv H(z)/H_0$ is the normalized Hubble parameter, which can be expressed as follows for a \wcdm\ cosmology,
\begin{equation}
    E(z)=\sqrt{ \Omega_{\rm m} (1+z)^3 +  \Omega_{\Lambda} (1+z)^{3(1+w)} }.
    \label{eq:E(z)}
\end{equation}
In \autoref{eq:E(z)}, \w\ represents the dark-energy equation-of-state parameter that indicates the scaling of dark-energy with time, and 
$\Omega_{\rm m} + \Omega_{\Lambda} = 1$.
Here, $w=-1$ gives the standard \lcdm \, cosmology.
In the expression of the distance ratio $\beta$, $H_0$ cancels out; thus, $\beta$ depends only on \om, \w,  and the redshifts of the deflector and sources.
Therefore, an independent measurement of redshifts and $\beta$ can constrain these cosmological parameters.

\section{Observations}
\label{Sec:data}

\begin{figure*}
    \centering
    \includegraphics[clip=true,trim= 02mm 0mm 0mm 0mm, width=  0.98\textwidth]{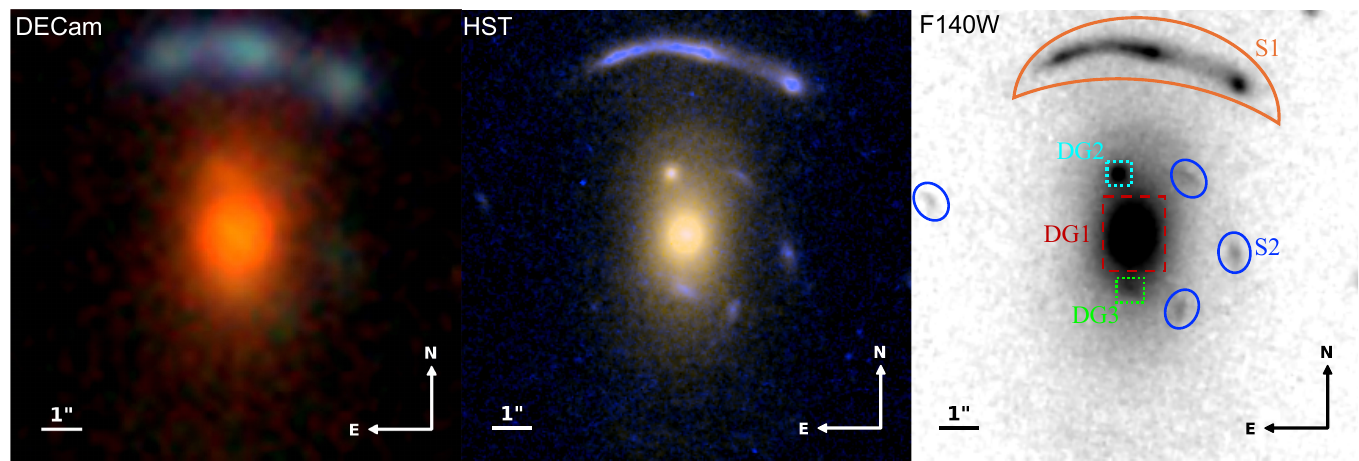}
    \caption{Ground-based and space-based images for DSPL \agel150745+052256.
    Left panel: DECaLS $grz$ color image from DECam observations. Middle panel: HST color image constructed using WFC3 F140W and F200LP filters. Right panel: HST/WFC3 F140W-band image with markers for the deflector galaxies (DG1, DG2, DG3) and the two background sources (S1 and S2).
    Image sides are $12\arcsec$ for each panel.
    Quadruply lensed images of the farthest source, S2 (\ztwo$=2.591$), are indicated by blue ellipses.
    The nearer source, S1 (\zone$=2.163$), has a naked-cusp configuration, i.e., it has three lensed images enclosed by the orange curve.
    Our spectroscopic observation and lens model confirm that the blue blob, labelled DG3 in the right-most panel, is not the counter-image of the S1 arc. 
    The main deflector galaxy, DG1 (\zdefl$=0.594$), is marked by a red square, and its satellite galaxies DG2 and DG3 are marked by cyan and green squares, respectively. 
    DG2 and DG3 are assumed to be at the same redshift \zdefl$=0.594$  as DG1.
    See \autoref{Sec:data} for further details on the observations and lens morphology.  
    }
    \label{fig:obs_img_des_hst}
\end{figure*}

For this work, we use the DSPL \agel150745+052256, shown in \autoref{fig:obs_img_des_hst}, from the \agel\ survey \citep{Tran:Harshan:2022}. 
It also has a Dark Energy Camera Legacy Survey \citep[DECaLS,][]{Dey:Schlegel:2019} catalog name DCLS1507+0522. 
The $grz$ color image for \agel1507 from DECaLS is presented in \autoref{fig:obs_img_des_hst}.
This system was first discovered as a single-source-plane lens, only detecting the bright arc of the first source north of the deflector galaxy,  in the DECaLS imaging using a convolutional neural network (CNN) based search for strong lenses by \citet{Jacobs:Collett:2019a, Jacobs:Collett:2019b}.
We identified a second, higher redshift background source 
using the Keck Cosmic Web Imager \citep[KCWI,][]{Morrissey:KCWI:2018} integral field spectroscopy as described below.
Subsequently, the high-resolution \textit{Hubble} Space Telescope (HST) images, shown in \autoref{fig:obs_img_des_hst}, revealed the faint arcs of the second source, confirming the double-source-plane lens detection.

\subsection{Spectroscopic observations}
\label{subsec: sepctroscopy} 
\agel1507 was observed with the KCWI on the Keck II telescope on 3 March 2022. Conditions were clear with a seeing of $0\farcs95$. Data were taken with the medium slicer using standard $2 \times 2$ binning, providing a $16\arcsec \times 20\farcs4$ field of view with $0\farcs3 \times 0\farcs7$ spatial pixels. We used the blue low-resolution grating resulting in spectral resolution $R \simeq 1800$ with coverage from $\simeq 3500-5500~\AA$. We obtained 7 exposures of 1200 seconds each at a PA of 90 degrees, for a total integration time of 140 minutes. The observed field of view covers all lensed images of both sources. Data were reduced following the same procedure as described in \cite{Vasan:Jones:2024}, to which we refer the readers for details. 

We extract the spectrum of the nearer source, Source 1 (S1), by summing the bright spaxels using an object mask. The resulting 1-D spectrum is shown in \autoref{fig:spectra}, revealing Ly$\alpha$ emission and many strong interstellar absorption lines typical of star-forming galaxies. We measure a redshift \zone$=2.163$ from the stellar photospheric absorption features. 
For the farther source, Source 2 (S2), we measure a redshift \ztwo$=2.591$ from the Ly$\alpha$ emission detected individually in the four images. The rest-frame integrated spectrum of S2 is shown in Figure \ref{fig:spectra src2}.
The deflector galaxy redshift was previously established as \zdefl$=0.594$ from the Baryon Oscillation Spectroscopic Survey of the Sloan Digital Sky Survey \citep[SDSS-BOSS, ][]{Eisenstein:Weinberg:2011, Dawson:Schlegel:2013}.

\subsubsection{Stellar Velocity Dispersion}
\label{subsubsec: velocity dispersion}

\begin{figure*}
    \centering
    \includegraphics[clip=true,trim= 0mm 0mm 0mm 0mm, width=  0.98\textwidth]{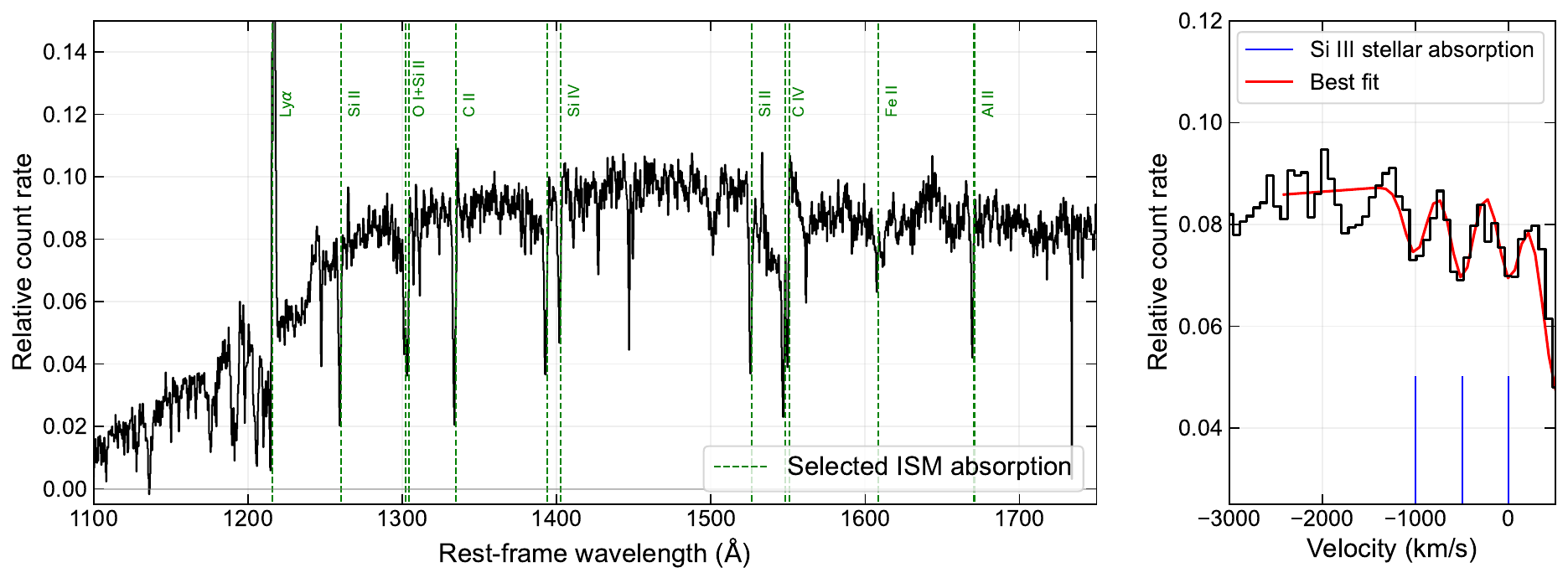}
    \caption{  Rest frame integrated spectrum for the Source 1 in the double-source-plane lens \agel1507 obtained using Keck/KCWI integral field spectroscopy (see \autoref{subsec: sepctroscopy}). We measure a redshift of \zone$=2.163$ for Source 1.
    Right panel shows the fit around the Si III stellar lines used to measure the velocity dispersion of Source 1 that is found to be $\sigma_{\rm obs}=109\pm27$\,km\,s$^{-1}$ (see \autoref{subsubsec: velocity dispersion}). }
    \label{fig:spectra}
\end{figure*}

\begin{figure}
    \centering
    \includegraphics[clip=true,trim= 02mm 0mm 0mm 6.8mm, width=  0.49\textwidth]{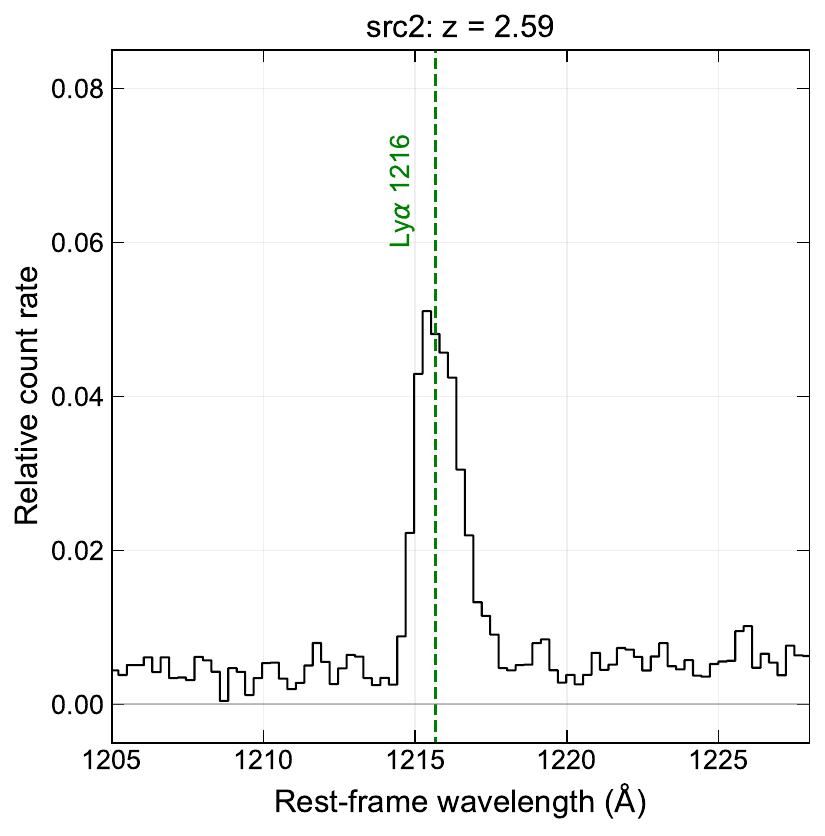}
    \caption{ Rest frame integrated spectrum for the Source 2 in DSPL \agel1507 obtained from the Keck/KCWI integral field spectroscopic data. We used the Ly$\alpha$ line to measure Source 2 redshift \ztwo$=2.591$ (see \autoref{subsec: sepctroscopy}).}
    \label{fig:spectra src2}
\end{figure}

The single-aperture line-of-sight stellar velocity dispersion of the central deflector galaxy in \agel1507 is measured by SDSS-BOSS survey to be $\sigma_{\rm obs}=303\pm38$~km~s$^{-1}$ \citep{Thomas:Steele:2013}.
 SDSS-BOSS survey used a circular fibre of diameter $2\arcsec$ for this observation.
We measure the velocity dispersion of the young stars in Source 1 from stellar photospheric lines in the KCWI spectrum. We use the relatively strong and unblended Si~III $\lambda1294$, Si~III $\lambda 1296$ and C~III $\lambda 1296$, and Si~III $\lambda \lambda 1298.89, 1298.96$ doublet complex. This complex, as well as the neighbouring interstellar O~I $\lambda 1302$ and Si~II $\lambda 1304$ lines, is simultaneously fit with Gaussian components and a linear continuum, enabling the full region around the stellar lines to be modeled. 
The best-fit to Si~III stellar absorption lines is shown in the right panel of \autoref{fig:spectra}.
The best-fit stellar velocity dispersion for Source 1 is measured to be $109 \pm 27$~km~s$^{-1}$.

\subsection{Imaging observations}
\label{subsec: imaging} 
The HST images for \agel1507 were taken by the Wide-Field Camera 3 (WFC3) in the IR/F140W filter via the SNAPSHOT program 16773 (Cycle 29-30, PI: K. Glazebrook).
Program 16773 followed a multifilter observation sequence  optimized using \textsc{LensingETC} \citep{Shajib:Glazebrook:Barone:2022} to be executed within one truncated HST orbit.
The lens was observed with the IR/F140W filter in three 200-second exposures and the UVIS/F200LP filter in two 300-second exposures.
Further observing details for HST images can be found in \agel\ data release 2 \citep{AGEL:DR2:2025}.
In \autoref{fig:obs_img_des_hst}, we present the $grz$ color image from DECaLS (left panel), the color image based on HST F140W and F200LP images (middle panel), and the F140W band image marking the deflector galaxies and background sources (right panel).

For lens modeling described in \autoref{Sec:modeling}, we use the higher wavelength image in the F140W filter because 
1) the light distribution traces the stellar mass distribution more closely in the longer wavelength, which is particularly useful for our work, where we perform composite modeling to constrain dark and stellar mass distributions individually,
2) the lensed images of the background sources appear less clumpy in this filter, allowing for an easier parametrized source reconstruction with fewer degrees of freedom needed during lens modeling.
The point spread function (PSF) of the image was produced using \textsc{Tiny Tim} \citep{Krist:2011}.

\subsection{Lens morphology}
\label{subsec: lensing morphology}

In \agel1507, there are three redshift planes: the deflector plane at \zdefl$=0.594$, the Source 1 plane at \zone$=2.163$, and the Source 2 plane at \ztwo$=2.591$.  
There are two satellite galaxies, DG2 and DG3, located north and south, respectively, of the main deflector galaxy (DG1), see the right panel in \autoref{fig:obs_img_des_hst}. 
Redshifts of DG2 and DG3 are unknown because they are faint, and 
we do not detect any identifying features in their KCWI spectra. 
For the reasons described in the next paragraph, we assume that the satellite galaxies DG2 and DG3 are at the same redshift as DG1.  
Thus, the deflector plane comprises the main galaxy DG1 and the satellite galaxies DG2 and DG3.  
S1 is lensed by the deflector plane galaxies DG1, DG2, and DG3.  
S2 is presumably lensed first by the S1 plane and then by the deflector plane, as depicted in \autoref{fig:ray_diag}.  
The deflector plane galaxies DG1, DG2, and DG3 are marked with red, cyan, and green squares, respectively, in \autoref{fig:obs_img_des_hst}.
The lensed images of the two background sources, S1 and S2, are shown by orange and blue contours, respectively, in \autoref{fig:obs_img_des_hst}.

\begin{figure*}
    \centering
    \includegraphics[clip=true,trim= 02mm 0mm 0mm 0mm, width=  \textwidth]{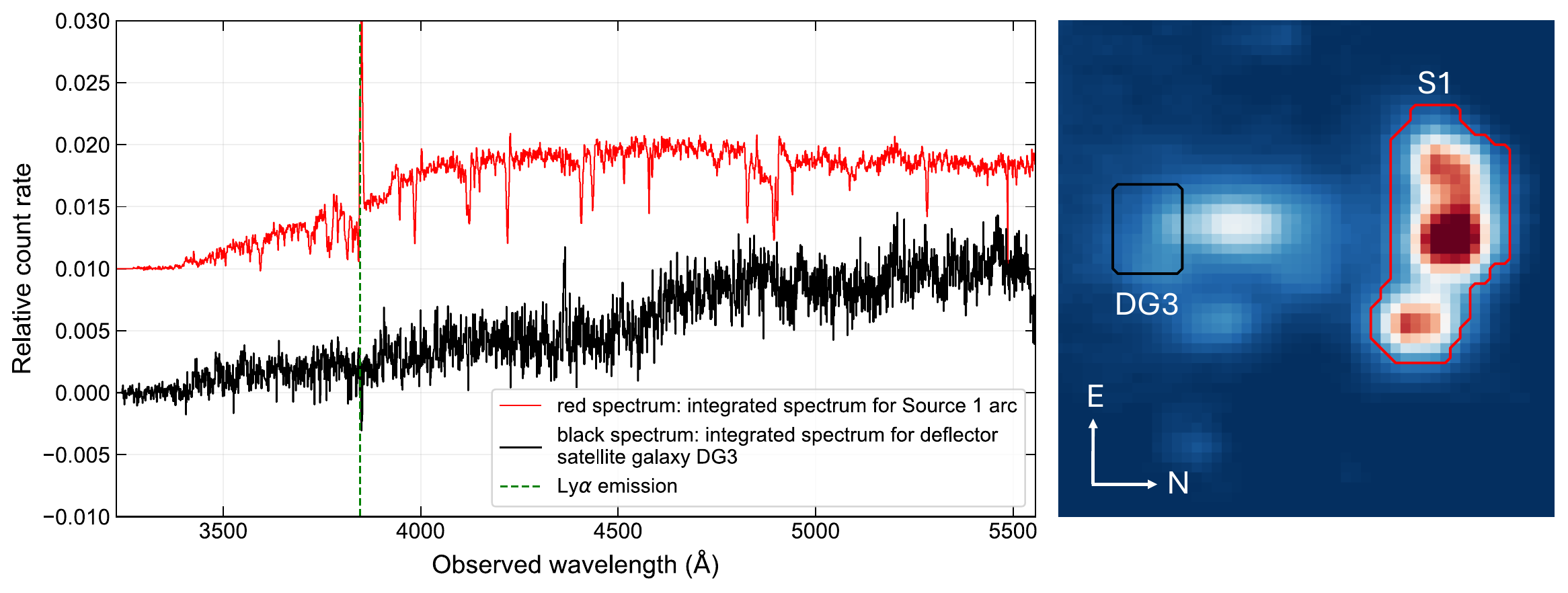}
    \caption{ Integrated spectra for Source 1 arc and deflector satellite galaxy DG3 (left panel) and the median KCWI map for \agel1507 (right panel). 
    The non-detection of Ly$\alpha$ emission in the DG3 region suggests it is not a part of Source 2 lensed image configuration (see \autoref{subsec: lensing morphology}). }
    \label{fig:kcwi map}
\end{figure*}

As seen in the HST color image in \autoref{fig:obs_img_des_hst}, the deflector satellite galaxy DG2 has a similar color to the main deflector galaxy DG1; therefore, we assume DG2 to be at the same redshift as DG1.
Satellite galaxy DG3 is located in the region where one could expect a counter-image of the S1 arc if it were a typical cross configuration.
However, we assume DG3 to be another perturber in the deflector plane, and not the counter-image, because:
i) the Ly$\alpha$ emission observed in the S1 arc is absent at the expected counter-image location, as shown in \autoref{fig:kcwi map},
ii) none of our models successfully reconstruct the counter-image of S1, particularly its orientation, indicating that the S1 arc is created by a naked-cusp \citep[i.e., three lensed images on one side of the deflector, see][]{Kochanek:2006}. 
In fact, the single-source-plane model for \agel1507 also did not suggest a counter image for S1 \citep[see Fig. A1 in ][]{Sahu:gamma_z:2024}.

\section{Lens Modeling} 
\label{Sec:modeling}

To model the DSPL \agel1507, we use the multi-purpose software package \textsc{lenstronomy} \citep{Birrer:Amara:2018, Birrer:2021}.
We employ the multi-plane ray-tracing implementation in \textsc{lenstronomy}, which allows freely varying the distance ratios.
First, we obtain a multi-plane lens model using fixed distances based on spectroscopic redshifts and a fiducial flat \lcdm\ cosmology with $H_{0} = 70 \,\rm km \, s^{-1} \, Mpc^{-1}$ and \om$=0.3$.
Once we find the best-fit model, we allow the distances to vary and sample the distance ratio, i.e., the cosmological scaling factor $\beta_{12}$.
The components of our lens model and the modeling process are described in the following sections.

\subsection{Lens model components}
\label{subsec:model}
We model the total mass of the main deflector galaxy DG1 using a composite model with two components: one for the dark matter halo and one for the baryonic matter.
For the dark matter halo component, we use an elliptical Navarro–Frenk–White \citep[NFW;][]{Navarro:Frenk:White:1997} convergence profile, approximated using a series of cored steep ellipsoids (CSE).
Readers are directed to \citet{Oguri:2021} for the expression of the effective convergence (i.e., dimensionless projected mass profile, $\kappa$) of the NFW profile.
The baryonic mass profile of DG1 is assumed to follow its light profile, with an additional stellar mass-to-light ratio ($M\/L$) parameter that is allowed to vary freely and converts light into stellar mass. We assume that all the baryonic mass in DG1 is in the form of stars.
For the light profile of DG1, we use a double Chameleon profile.

A Chameleon profile is the difference between two power-law profiles which approximates a S\'ersic profile within $1-2\%$  over the radial range of 0.5 to 3 times the half-mass radius for S\'ersic indices between 1 and 4 \citep[see][]{Dutton:Brewer:2011}.
Although the family of S\'ersic profiles \citep{Sersic:1963, Sersic:1968} is known to describe the light profiles of galaxies very well, calculating lensing quantities based on the S\'ersic profile is complex and computationally expensive.
In contrast, the Chameleon profile has a simpler analytical expression, allowing for faster computation of lensing quantities \citep[see][for the expression of the Chameleon convergence profile]{Suyu:Treu:2014, Shajib:2019}.

For the deflector satellite galaxy DG2, we use a singular isothermal sphere (SIS) model for the mass and a circular S\'ersic model for the light profile. For the deflector satellite galaxy DG3, we use a singular isothermal ellipsoid (SIE) model for the mass and an elliptical S\'ersic model for the light profile. These profiles, which include additional ellipticity parameters, are chosen to accommodate the high ellipticity of DG3. In addition, we include a residual shear (i.e., external shear) component in the deflector plane to account for the remaining tidal lensing potential from the deflector’s local environment and possible complexity in the central deflector's angular structure \citep{Etherington:2024}.

For Source 1, we use an SIE model for the mass and model the surface brightness distribution using an elliptical S\'ersic profile combined with a basis set of shapelets \citep{Refregier:2003, Refregier:Bacon:2003, Birrer:Amara:2015}.
The surface brightness distribution for Source 2 is reconstructed using an elliptical S\'ersic profile.
For more details about parameterization and the convergence profile, readers are directed to the latest \textsc{lenstronomy} \href{https://lenstronomy.readthedocs.io/}{documentation}.

\subsection{Modeling process}
\label{subsec:modeling process}

We perform the lens modeling in two phases. First, we construct a preliminary mass model using only the positions of the lensed images for both sources relative to the deflector as constraints when solving the lens equation (\autoref{eq:lens eq multi-plane}).
Second, we refine this model through extended modeling, incorporating the full lensing information -- specifically, the surface brightness distribution of the lensed arcs and images -- as additional constraints.

During extended modeling, we use the deflector mass model parameters derived from the position modeling as initial values.
To obtain the parameters of the double Chameleon light profile of DG1, we separately fit a double Chameleon model to a double S\'ersic profile that describes its light distribution down to the noise level.
Additionally, we align the centers of the dark matter and baryonic matter components of DG1.
For the initial light profile parameters of DG2, DG3, and the background sources S1 and S2, we make a reasonable estimate and align the mass profile centers of DG2, DG3, and S1 with their corresponding photometric centers.
For S1, we also enforce alignment between the ellipticity parameters of the mass and light profiles to reduce the uncertainty in its mass model and decrease the number of free parameters to be sampled. For S1 shapelet profile, we fix the maximum polynomial order of the shapelet basis set to 8. Finally, we use uniform priors with wide bounds for all model parameters.

During position modeling, we use \textsc{scipy} optimization to obtain a preliminary mass model.
For extended modeling, we first apply Particle Swarm Optimization \citep[PSO,][]{Kennedy:1995}, followed by Markov Chain Monte Carlo (MCMC) sampling using \textsc{emcee} \citep{Foreman-Mackey:Hogg:2013} until convergence is reached.
We consider the chain to have converged when the median and standard deviation of the \textsc{emcee} walkers remain in equilibrium for at least the last 1,000 steps. The initial PSO optimization helps rapidly approach the maximum of the posterior distribution by providing a starting point for the MCMC that is closer to the maximum than an otherwise arbitrary starting point.
Subsequently, MCMC sampling explores the region around the best solution found by PSO, providing the posterior probability distribution of the model parameters.

First, we perform a complete lens modeling using our fiducial cosmological model, as mentioned earlier.
Once the best-fit lens model is obtained upon the convergence of the MCMC chain, we set the distance ratio $\beta_{12}$ (see section \ref{subsec: Beta}) free and sample its posterior distribution along with other lens model parameters using MCMC until the convergence is achieved again.
We use a uniform prior for $\beta_{12}$, with its range determined based on a uniform range for \om$\in[0,1]$ and $w\in[-2,0]$.

\begin{figure*}
    \centering
    \includegraphics[clip=true,trim= 60mm 20mm 10mm 20mm, width=1.1\textwidth]{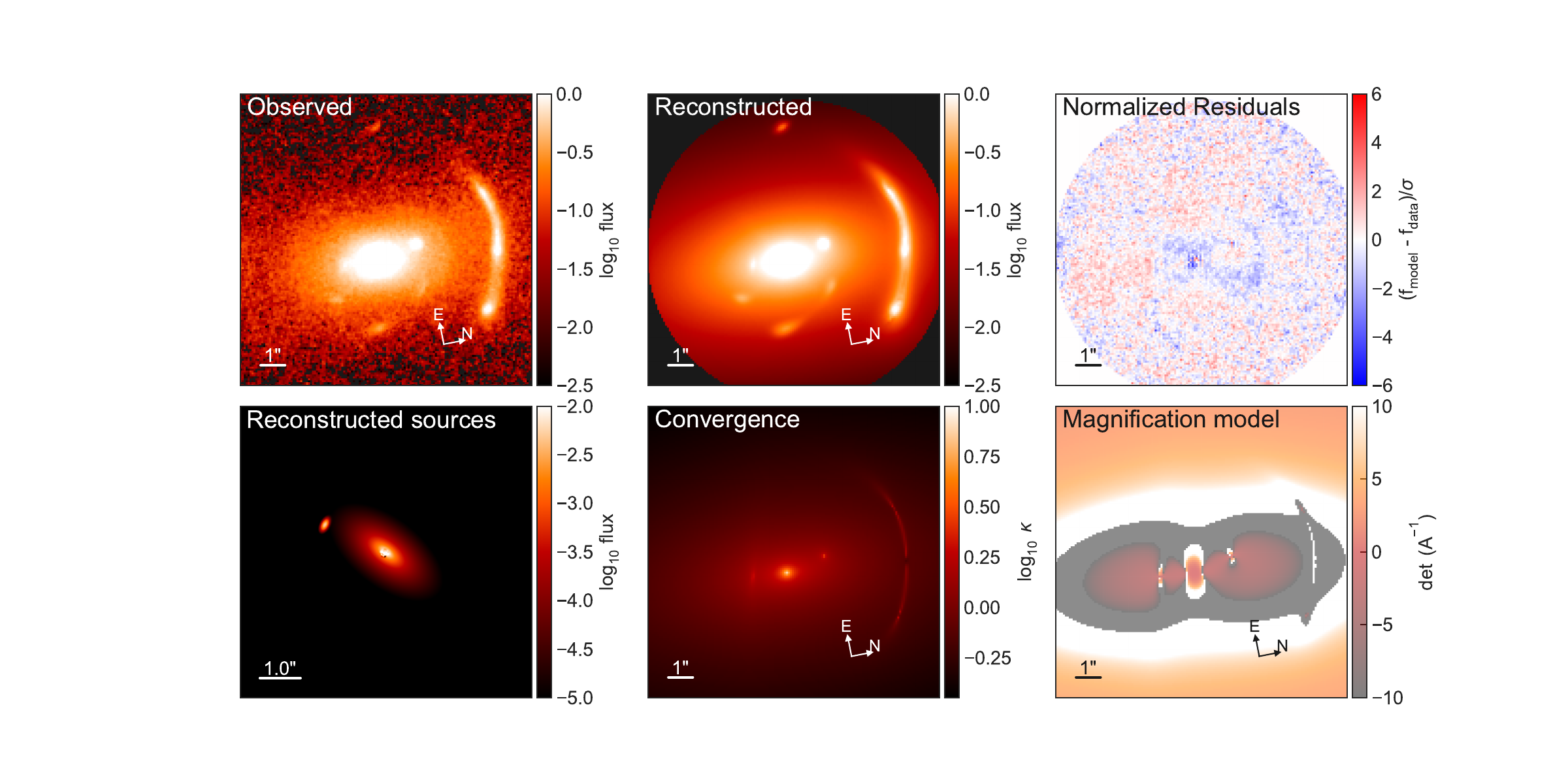}
    \caption{Best-fit model for \agel1507 modeled with \textsc{lenstronomy} in multi-plane mode using the HST/WFC3 F140W band image.
    The top panels show, from left to right, the observed lens image, the reconstructed lens model, and the normalized residual.
    Our lens model successfully reconstructs the naked-cusp (three lensed image) configuration for S1 and quadruply lensed images for S2 (see \autoref{subsec: lensing morphology}). 
    The surface brightness distributions for the main deflector and the two nearby satellite galaxies are also modeled well, as apparent from the residual.
    The bottom left panel displays the reconstructed images of the two sources. In contrast, the bottom middle and right panels show the effective convergence and the effective magnification map, respectively, for S2 projected onto the main deflector plane (\zdefl$=0.594$). 
    The effective convergence (and the corresponding critical curve in the `Magnification model') due to S1 (\zone$=2.163$) has an arc shape after being projected onto the main deflector plane.
    Various physical properties of the deflector components in DSPL \agel1507 based on our final lens model are summarized in \autoref{table:deflector properties}. 
    See \autoref{Sec:modeling} and \autoref{Sec:modeling results} for the modeling process and results. }
    \label{fig:model_composite_F140W}
\end{figure*}

\section{Modeling results}
\label{Sec:modeling results}

Following the modeling process described above, we obtain the best-fit lens model for \agel1507 using the HST F140W band image, as shown in \autoref{fig:model_composite_F140W}.
The top panels of \autoref{fig:model_composite_F140W} display, from left to right, the observed lens image, the reconstructed model, and the normalized residual.
The bottom panels show the reconstructed images of the two sources, the effective convergence, and the effective magnification map on the deflector plane (\zdefl$=0.594$) with respect to the farther source.

\begin{figure*}
    \centering
    \includegraphics[clip=true,trim= 0mm 10mm 0mm 10mm, width=\textwidth]{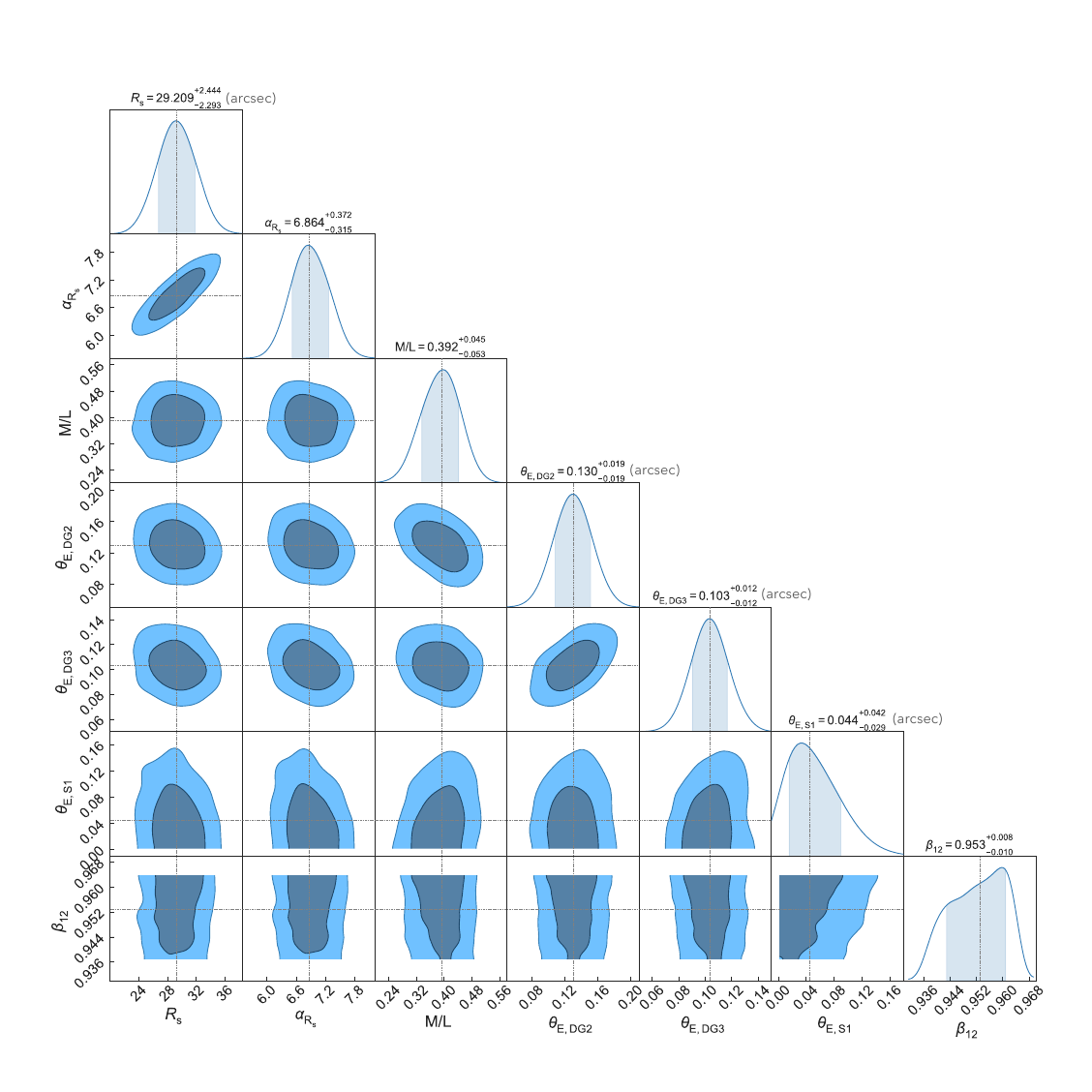}
    \caption{2D distribution of deflector mass profile parameters and the distance ratio factor $\beta_{12}$ from the final converged MCMC chain (see \autoref{subsec:modeling process}). Inner and outer contours represent $68\%$ and $95\%$ credible regions about the median of the posterior distribution. Our modeling results are described in \autoref{Sec:modeling results}, demonstrated in \autoref{fig:model_composite_F140W}, and summarized in \autoref{table:deflector properties}. 
    }
    \label{fig:corner plot}
\end{figure*}

\begin{table*}
    \centering
    \begin{tabular}{|l|c|c|c|c|c|c|} 
    \hline
    \textbf{Component \Big / Property} & \textbf{Total mass} & \textbf{$\theta_{\rm E}$} & 
    \textbf{Magnitude} &  \textbf{$R_{\rm half}$}& \textbf{$\sigma_{\rm pred}$} & \textbf{$\sigma_{\rm obs}$} \\ 
    %\hline
                        & $\log_{10} \ (M/M_\odot)$ & (arcsec) & (F140W) & (arcsec) & (km/s) & (km/s)\\ 
    \hline
    \textbf{DG1 (total)}&12.78$\pm$0.02 & $3.10^{+0.08}_{-0.04}$ & - & - & 290 $\pm$30 & 303$\pm$38 \\  
    %\hline
    \textbf{DG1 (stars)}&11.45$\pm$0.06&  - &  17.79$\pm$0.01 & 1.94 $\pm$0.06 & -  & - \\   
    %\hline
    \textbf{DG2}&10.49$\pm $0.07 &0.13$\pm$0.02 &  21.50$\pm$0.24  &  0.32$\pm$0.03  & 69$\pm$5 & -  \\ 
    %\hline
    \textbf{DG3}&9.81$\pm$ 0.06 & 0.10$\pm$0.01 &  23.38$\pm$0.40  & 0.14$\pm$0.02 &  52$\pm$6   & -  \\ 
    %\hline
    \textbf{S1 (intrinsic)}& $10.77^{+0.28}_{-0.51}$ & $0.04^{+0.04}_{-0.03}$ &   23.85$\pm$0.10   & 0.19$\pm$0.01 & 97$\pm$33 & 109$\pm$27 \\ 
    \hline
    \end{tabular}
    \caption{
    Mass, Einstein radii, apparent magnitudes (in the AB system), half-light radii, and the predicted and observed stellar velocity dispersions for deflector components in the DSPL \agel1507.  
    Total mass is measured within three half-light radii of each galaxy.  
    For these measurements, we use the final \wcdm\ cosmology, obtained by combining our results with CMB constraints, as presented in section \ref{subsec: cosmology conbined with other probes}. 
    For the main deflector galaxy DG1, $\theta_{\rm E}$ represents the effective projected radius enclosing an average convergence of 1, calculated using its dark matter (elliptical NFW) and baryonic matter (double chameleon) components.  
    Parameters for S1 are intrinsic, i.e., magnification corrected. 
    For all deflector components, $R_{\rm half}$ represents the circularized half-light radius, defined as the geometric mean of the half-light radii along the major and minor axes. 
    See \autoref{Sec:modeling results} for detailed modeling results.
     }
    \label{table:deflector properties}
\end{table*}

A corner plot presenting the 2D posterior distributions of the primary mass model parameters for the deflector components is shown in \autoref{fig:corner plot}.  
These parameters include the angular scale radius ($R_{\rm s}$) of the NFW profile, the deflection angle at the scale radius ($\alpha_{\rm R_s}$), and the mass-to-light ratio ($M/L$) for the baryonic component of the main deflector galaxy DG1.  
Additionally, the plot includes the Einstein radius of satellite galaxy DG2 ($\theta_{\rm E, DG2}$), the Einstein radius of satellite galaxy DG3 ($\theta_{\rm E, DG3}$), the Einstein radius of Source 1 ($\theta_{\rm E, S1}$), and the cosmological distance ratio parameter ($\beta_{12}$).
All parameters for the deflector mass profiles, deflector light profiles, and reconstructed source profiles are provided in the appendix \autoref{table:model parameters}. 

Physical properties such as the total mass, Einstein radii, apparent magnitudes, half-light radii, and the predicted and observed velocity dispersions of the deflector components are summarized in \autoref{table:deflector properties}.  
Unless stated otherwise, we report the median value along with the $68\%$ credible region for all parameters.
For cosmology-dependent quantities presented here -- such as mass, velocity dispersion, and halo concentration -- we adopt a flat \wcdm\ cosmology, obtained by combining DSPL constraints with those from the \textit{Planck} CMB observations, as presented in the next section. 
Since our cosmology constraints are independent of \( H_{0} \), we assume $H_{0} = 70 \,\rm km \, s^{-1} \, Mpc^{-1}$.

Our main modeling results are the following:
\begin{itemize}
    \item The main deflector galaxy DG1 (\(z = 0.594\)), modeled using an elliptical NFW profile for the dark matter component and a double Chameleon profile for the stellar mass component, has an effective projected Einstein radius of \(3\farcs10^{+0.08}_{-0.04}\) for Source 2.  
    Within three half-light radii of the deflector galaxy, the total mass (dark matter plus stellar) is found to be $\log_{10}(M/M_\odot) = 12.78 \pm 0.02$, while the stellar mass is $\log_{10}(M_{\star}/M_\odot) = 11.45 \pm 0.06$.

    \item Our model-predicted velocity dispersion for the main deflector galaxy is $290 \pm 30$ km\,s$^{-1}$, which is consistent with the directly measured value of $303 \pm 38$ km\,s$^{-1}$ from SDSS-BOSS observations, within the $1\sigma$ uncertainty bound.
    
    \item Satellite deflector galaxy DG2 ($z=0.594$) has an Einstein radius of $0\farcs13 \pm 0\farcs02$ and a total mass of $\log_{10}(M/M_\odot) = 10.49\pm0.07 $  within its three half-light radii. For this galaxy, our model predicts a velocity dispersion of $69 \pm 5$ km\,s$^{-1}$.

    \item Satellite deflector galaxy DG3 ($z=0.594$) has an Einstein radius of $0\farcs10 \pm 0\farcs01$, a total mass of $\log_{10}(M/M_\odot) = 9.81\pm 0.06$ within its three half-light radii, and a model-predicted velocity dispersion of $52 \pm 6$ km\,s$^{-1}$.

    \item Source 1 ($z=2.163$) has an Einstein radius of $0\farcs05^{+0.04}_{-0.03}$ and a total mass of $\log_{10}(M/M_\odot) =10.77^{+0.28}_{-0.51}$ within its three half-light radii. Our model predicts a velocity dispersion of $97 \pm 33$ $\rm km \, s^{-1}$, which is consistent with our directly measured velocity dispersion of $109 \pm 27$ km\,s$^{-1}$ within the $1\sigma$ uncertainty bound.

    \item The median value of the cosmological scaling factor is $\beta_{12} = 0.953^{+0.008}_{-0.010}$.  
    
\end{itemize}

Velocity dispersion predictions are obtained using the \textsc{Galkin} module in \textsc{lenstronomy} via spherical Jeans modeling. The anisotropy of stellar orbits is a key parameter that affects velocity dispersion estimates \citep[see][]{Birrer:Shajib:Galan:2020}.  
For our predictions, we marginalize over the impact of anisotropy by uniformly varying (i.e., imposing a uniform prior on) the anisotropy scale radius in the Osipkov--Merritt anisotropy profile between 0.5 and 5 times the half-light radius \citep{Osipkov:1979, Merritt:1985}.

When calculating the lens model-based velocity dispersion for the main deflector galaxy DG1, we used the same observational setting as for $\sigma_{\rm obs}$ from SDSS-BOSS, namely, a circular aperture with a diameter of $2\arcsec$ and a seeing disk full width at half maximum (FWHM) of $1\farcs0$.  
The model-predicted velocity dispersion for Source 1 is calculated using a circular aperture of radius twice the half-light radius and a seeing FWHM of $0.95\arcsec$ as per our Keck/KCWI observation.  
For the satellite galaxies DG2 and DG3, we assumed the same KCWI observational settings for calculating $\sigma_{\rm pred}$.

The dark halo component of the main deflector galaxy DG1 has a virial radius of $r_{200} = 0.93 \pm 0.03 \ \rm Mpc$, a virial mass of $\log_{10}(M_{200}/M_\odot) = 14.21 \pm 0.04$, and a halo concentration of $c_{200} = r_{200}/R_{s} = 4.63 \pm 0.22$. 
For the same halo mass, redshift, and cosmology, the semi-analytical dark matter halo evolution model from \citet{Diemer:Joyce:2019} predicts $c_{200} = 3.8 \pm 0.2$ with a scatter of $\pm 1.4$.  
Although our lens model-based median value of $c_{200}$ is higher than that from \citet{Diemer:Joyce:2019}, it remains consistent within $3\sigma$ of their median value and lies well within the $1\sigma$ scatter.  
The stellar-to-halo mass ratio, $\log_{10}( M_{\star}/M_{200}) = -2.76 \pm 0.07$, is consistent within the $1\sigma$ bound of the empirical stellar-to-halo mass relation in \citet{Behroozi:2019} and \citet{Girelli:2020}.

The satellite galaxies DG2 and DG3, which we assume to be at the same redshift as the main deflector DG1, have a very small lensing contribution, as expected from their small Einstein radii. Their total mass accounts for only $\sim 0.6\%$ of the total deflector mass, suggesting minimal effect on the overall lensing configuration. Therefore, our assumption of treating DG2 and DG3 at the same redshift as DG1 does not significantly impact our modeling results.

\citet{Schneider:2014} pointed out the impact of mass-sheet degeneracy (MSD) with multiplane lensing for studying cosmology through DSPL.  The MSD is inherent to lensing, and breaking the MSD requires additional information.  The kinematic measurements we have obtained on the deflectors and the source S1 provide such additional information, since a mass-sheet transformation would change the predicted kinematics of the deflectors and S1.  The good agreement we have obtained in the predicted and measured velocity dispersion of S1 is reassuring and helps limit the effect of MSD.  We defer to future studies for a more thorough and joint analysis of lensing and kinematic data for breaking the MSD.

\section{Cosmological constraints}
\label{Sec:Comological inferences}

\begin{figure}
    \centering
    \includegraphics[clip=true,trim= 00mm 01mm 02mm 10mm, width=0.5\textwidth]{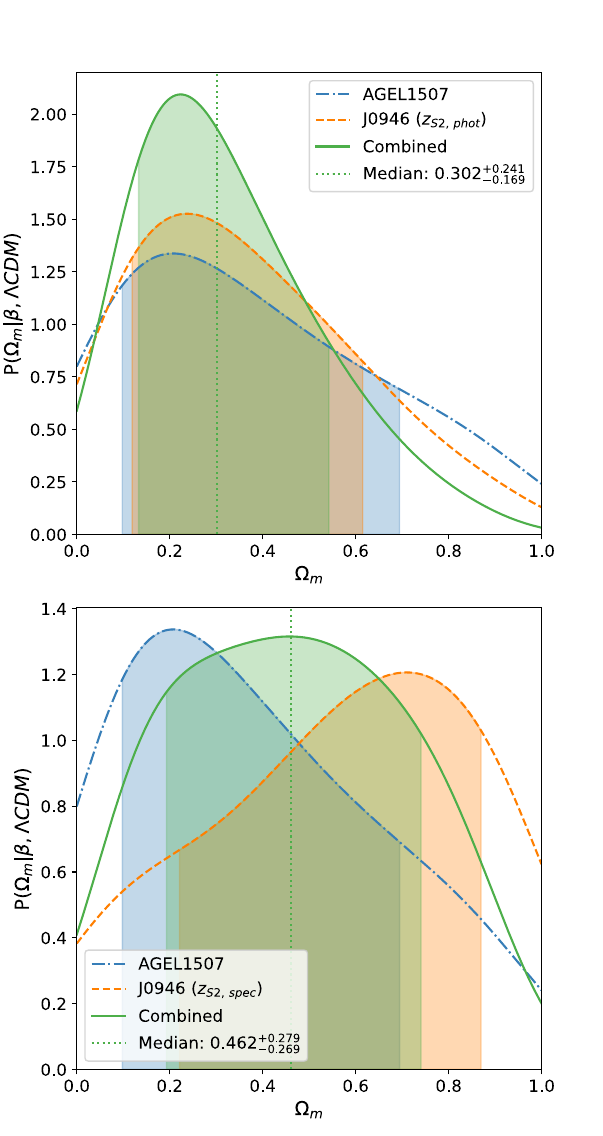}
    \caption{Probability distribution function (PDF) of \om \ in \lcdm \ cosmology given $\beta$ obtained from lens modeling. The individual distributions for \agel1507 and J0946 are shown by the dot-dashed blue and dashed orange curves, respectively. The \om\ PDF for \agel1507 is the same in both panels (see \autoref{subsec:agel1507}). 
    In the top panel, the PDF of \om\ for J0946 is obtained using the photometric redshift ($z_{\scalebox{0.5}{S2}, \, \rm phot}=2.4$) for the farther Source 2  \citep[as used in][]{Collett:Auger:2014}, see \autoref{subsec:J0946_phot_z_s2}.
    The bottom panel shows \om\ distribution obtained using new spectroscopic redshift ($z_{\scalebox{0.5}{S2}, \, \rm spec}=2.035$) for Source 2 in J0946 (see \autoref{subsec:J0946_spec_z_s2}).  The green curve shows the combined PDF of \om\ in each panel. 
    The shaded region represents the $68\%$ credible interval.
    The final constraints after combining those from \agel1507 and updated J0946 are discussed in \autoref{subsec:combined_DSPLs} and summarized in \autoref{tab:cosmology}.
    }
    \label{fig:Om_LCDM}
\end{figure}

Using the posterior distribution of the cosmological scaling factor $\beta_{12}$ and independently measured spectroscopic redshifts for the deflector and two background sources, we derive constraints on the following cosmological models.\\
\textbf{i) Flat \lcdm\ model}: Assuming a flat Universe (\ok$=0$) with a constant dark-energy equation-of-state parameter ($w=-1$), 
$\beta_{12}$ depends only on the matter density parameter \om\ and the redshifts of the deflector and sources. 
Using $\beta_{12}$ obtained from our lens model and measured redshifts, we constrain \om.\\
\textbf{i) Flat \wcdm\ model}: Assuming a flat Universe (\ok$=0$),  $\beta_{12}$ is a function of \om, $w$, and the redshifts. By utilizing $\beta_{12}$ and the redshift measurements, we obtain joint constraints on \om\ and $w$.\\
In all cases, we adopt uniform priors for the cosmological parameters: \om$\sim \mathcal{U}[0,1]$ and $w \sim \mathcal{U}[-2,0]$.
\autoref{tab:cosmology} summarizes all the inferences presented in this section.

\subsection{Constraints from \agel1507} 
\label{subsec:agel1507}

Based on $\beta_{12}$ obtained from the modeling of DSPL \agel1507 and spectroscopically measured redshifts for the deflector and two background sources, we infer a median value of \om$= 0.334^{+0.380}_{-0.234}$ for the flat \lcdm\ model. 
The error bars represent the $68\%$ credible bounds around the median.
The full probability distribution function (PDF) of \om\ is shown by the dot-dashed blue curve in \autoref{fig:Om_LCDM}, with the filled regions indicating the $68\%$ credible interval around the median.

For a flat \wcdm\ model, we find \om$= 0.344^{+0.384}_{-0.239}$ and $w= -1.24^{+0.66}_{-0.51}$ based on our measurement of $\beta_{12}$ from \agel1507. The 2D distribution of \om\ and $w$ for \agel1507 is shown by the blue region in the left panel of \autoref{fig:wCDM}, where the dark and light shaded regions indicate the $68\%$ ($1\sigma$) and $95\%$ ($2\sigma$) credible regions.

\subsection{Constraints from J0946} 
\label{subsec:J0946_phot_z_s2}

To constrain cosmology using their DSPL model for J0946, \citet{Collett:Auger:2014} used spectroscopic redshifts for the deflector (\zdefl$=0.222$) and the nearer source (\zone$=0.609$);  however, they used a photometric redshift for the farther source ($z_{\scalebox{0.5}{S2}, \, \rm phot}\sim 2.4$).  
From the lens modeling, they found $\beta_{12}=0.712 \pm 0.008$ for J0946 and derived constraints on \om\ and $w$,  which were marginalized over the photometric redshift of the farther source.  
They inferred \om$= 0.33^{+0.33}_{-0.26}$ for the \lcdm\ model and, when combined with CMB constraints, inferred $w = -1.17^{+0.20}_{-0.21}$ for the flat \wcdm\ model.  
The full PDF of \om\ for the \lcdm\ model, obtained using a redshift of $z_{\scalebox{0.5}{S2}, \, \rm phot}=2.4$ for the second source (taken from their Fig. 6), is shown by the dashed orange curve in the top panel of \autoref{fig:Om_LCDM}. The inferred \om\ and \wo\ parameters for \lcdm\ and \wcdm\ models using $z_{\scalebox{0.5}{S2}, \, \rm phot}=2.4$ are presented in \autoref{tab:cosmology}.

\subsection{Updated constraints from J0946} 
\label{subsec:J0946_spec_z_s2}

\citet{Smith:Collett:2021} later confirmed the spectroscopic redshift of the second source to be $z_{\scalebox{0.5}{S2}, \, \rm spec}=2.035$ using Very Large Telescope X-shooter observations and reported the updated DSPL plus CMB constraints for the flat \wcdm\ model to be $w= -1.04\pm 0.20$.  
Using the new spectroscopic redshift for the second source in J0946 and $\beta_{12}$ from \citet{Collett:Auger:2014}, we obtain \om$= 0.597^{+0.269}_{-0.383}$ for the \lcdm\ model  
and \om$= 0.621^{+0.257}_{-0.386}$ and $w= -0.76^{+0.51}_{-0.99}$ for the \wcdm\ model from J0946 alone.  
The full distribution of the updated constraints on \lcdm\ and \wcdm\ cosmologies from J0946 is presented in the bottom panel of \autoref{fig:Om_LCDM} (orange dashed curve) and the middle panel of \autoref{fig:wCDM} (orange region), respectively.

As notable from the two panels of \autoref{fig:Om_LCDM}, the updated \om\ posterior from J0946 based on the new spectroscopic redshift ($z_{\scalebox{0.5}{S2}, \, \rm spec} = 2.035$) for the second source, has shifted by approximately $1\sigma$ and exhibit a greater uncertainty compared to the constraints derived from the higher photometric redshift ($z_{\scalebox{0.5}{S2}, \, \rm phot} \sim 2.4$) used in \citet{Collett:Auger:2014}.  
A similar posterior shift and enhanced uncertainty are also seen for the \wcdm\ constraints from J0946 (see \autoref{tab:cosmology}). 
The enhanced uncertainty is qualitatively consistent with the expected increase in uncertainty as the redshift gap between sources decreases \citep[see][their Fig. 5]{Collett:Auger:2012}.  
This dramatic change in the inferred cosmology highlights the crucial role of spectroscopic confirmation in ensuring robust cosmological inference.

\citet{Collett:Smith:2020} additionally discovered a third lensed source at a redshift of $z=5.975$ in J0946 using Multi-Unit Spectroscopic Explorer (MUSE) observations.  
\citet{Ballard:Enzi:2024} recently updated the lens model for J0946, incorporating the lensed positions of the third source to constrain the lens model.  
An updated sampling of additional distance ratio parameters ($\beta_{13}, \beta_{23}$) for a triple-source-plane lens and cosmological inference is currently in preparation (Ballard et al., in preparation).

\subsection{Combined Constraints from DSPLs} 
\label{subsec:combined_DSPLs}

\begin{figure*}
    \centering
    \includegraphics[width=1\textwidth]{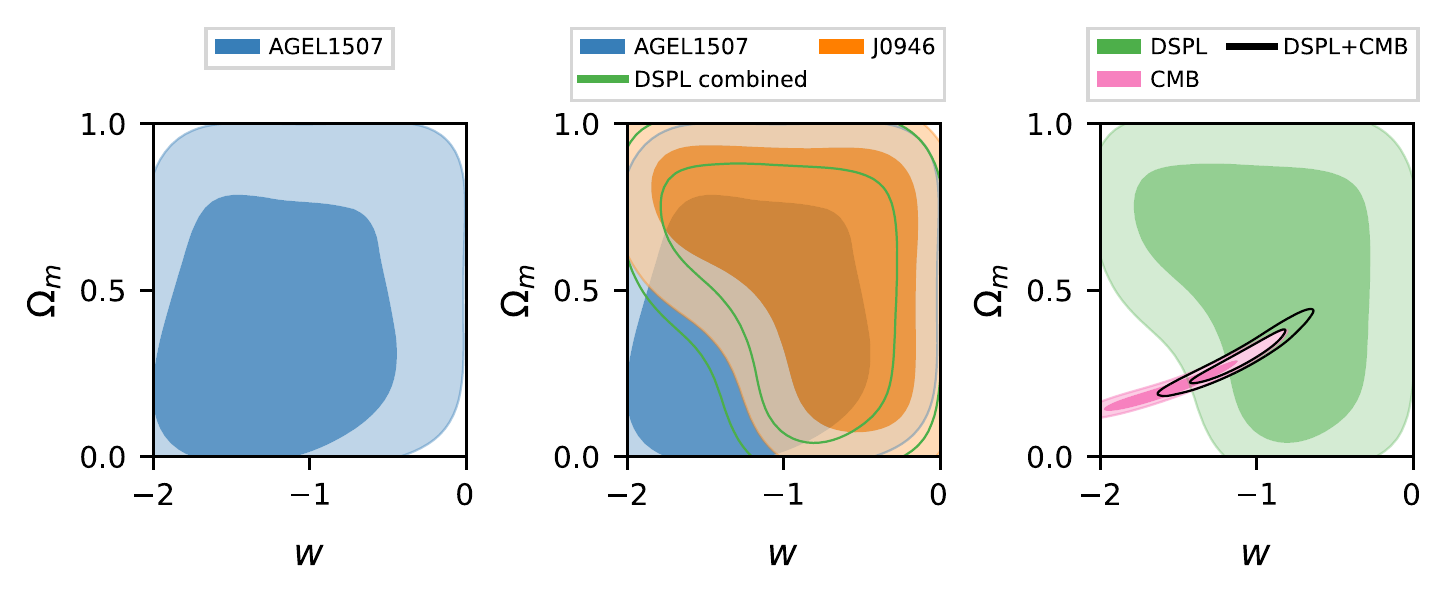}
    \caption{ 2D distribution of \wo\ and \om \ for the flat \wcdm\  model of the cosmology.
    Independent constraints from \agel1507, J0946, and  CMB data are represented by blue, orange, and pink regions, respectively. Individual constraints from \agel1507, J0946, and  CMB data are described in \autoref{subsec:agel1507}, \autoref{subsec:J0946_spec_z_s2}, and \autoref{subsec: cosmology conbined with other probes}.
    The green contours in the middle panel show combined constraints from DSPLs \agel1507 and J0946. The black contours present the final combined constraints from DSPLs and CMB data, which are discussed in \autoref{subsec: cosmology conbined with other probes}. 
    Dark (inner contour) and light (outer contour) shaded regions represent $68\%$ and $95\%$ credible regions.
    Individual and combined constraints are also summarized in \autoref{tab:cosmology}. }
    \label{fig:wCDM}
\end{figure*}

To obtain the combined constraints from independent observations, the probability distribution functions $P_{i}(x)$ from each independent observation $i$ were multiplied. 
The resulting product was normalized following \autoref{eq:combined pdf} to ensure that the combined PDF, $P_{\rm combined}(x)$, integrates to 1 

\begin{equation}
    P_{\rm combined}(x)=\frac{\prod_{i} P_{i}(x) }{\int \prod_{i} P_{i}(x)\ dx}.
    \label{eq:combined pdf}
\end{equation}

Combining the constraints from the two DSPLs, \agel1507 and J0946 (with the updated spectroscopic redshift for the second source), yields \om$ = 0.462^{+0.279}_{-0.269}$ for the \lcdm\ model.  
The precision of the joint \om\ measurement is improved by 10\% and 15\% compared to that from \agel1507 and J0946 individually.  
The PDF of the joint \om\ is shown by the green curve in the bottom panel of \autoref{fig:Om_LCDM}, where the filled region represents the 68\% credible interval around the joint median \om.  

For the \wcdm\ model, we find \om$= 0.545^{+0.273}_{-0.343}$ and $w = -0.89^{+0.49}_{-0.57}$.  
The joint constraints on \om\ and $w$ have 4\% and 30\% higher precision, respectively, compared to J0946 alone, and 1\% and 10\% higher precision compared to \agel1507 alone.  
The green contours in the middle and left panels of \autoref{fig:wCDM} represent the joint distribution of \om\ and $w$, obtained by combining the constraints from \agel1507 and J0946.  
The inner and outer green contours in \autoref{fig:wCDM} enclose the 68\% and 95\% credible regions, respectively.

Increasing the sample from one to two significantly improves the inference, especially for \wo, demonstrating substantial potential for stringent constraints from a larger sample consisting solely of DSPLs.
\autoref{tab:cosmology} presents a summary of individual and combined constraints from the two DSPLs and joint constraints with other probes of cosmology discussed next.

\begin{table}
\centering
\renewcommand{\arraystretch}{1.2} 
\begin{tabular}{|l|c|c|}
\hline
\multicolumn{3}{c}{\textbf{Flat \lcdm} \, ( \ok$=0$, $w=-1$) }  \\
\hline
     & \multicolumn{2}{|c|}{\textbf{\om } }       \\ \hline
\textbf{\agel1507}            & \multicolumn{2}{|c|}{

$ 0.334^{+0.380}_{-0.234} $  }        \\ \hline
\textbf{J0946} ($z_{\scalebox{0.5}{S2, phot}}$) & \multicolumn{2}{|c|}{$ 0.324^{+0.279}_{-0.206} $  }       \\ \hline
\textbf{J0946}   ($z_{\scalebox{0.5}{S2, spec}}$, updated) & \multicolumn{2}{|c|}{ $ 0.597^{+0.269}_{-0.383} $ }         \\ \hline
\textbf{DSPLs} (\agel1507 $+$ J0946) & \multicolumn{2}{|c|}{$ 0.462^{+0.279}_{-0.269} $ }           \\ \hline
\multicolumn{3}{c}{\textbf{Flat \wcdm} \,(\ok$=0$) }  \\ \hline
       & \textbf{\om}        & \textbf{\wo}          \\ \hline
\textbf{\agel1507} & $ 0.344^{+0.384}_{-0.239} $  &   $ -1.24^{+0.66}_{-0.51} $    \\ \hline
\textbf{J0946} ($z_{\scalebox{0.5}{S2, phot}}$ ) & $ 0.451^{+0.238}_{-0.253} $ &   $ -1.25^{+0.42}_{-0.51} $  \\ \hline
\textbf{J0946}   ($z_{\scalebox{0.5}{S2, spec}}$, updated) & $ 0.621^{+0.257}_{-0.386} $  & $ -0.76^{+0.51}_{-0.99} $  \\ \hline
\textbf{DSPLs} (\agel1507 $+$ J0946) & $ 0.545^{+0.273}_{-0.343} $  &   $ -0.89^{+0.49}_{-0.57} $      \\ \hline
\textbf{CMB} & $ 0.198^{+0.068}_{-0.038} $  & $ -1.57^{+0.35}_{-0.27} $    \\ \hline
\textbf{DSPL $+$ CMB} & $ 0.293^{+0.060}_{-0.050} $  & $ -1.10^{+0.20}_{-0.22} $    \\ \hline
%\textbf{CMB $+$ SNe $+$ BAO} & $ 0.316^{+0.009}_{-0.008} $  & $ -0.96^{+0.03}_{-0.03} $    \\ \hline
\textbf{DSPL $+$ CMB $+$ SNe $+$ BAO} & $ 0.316^{+0.009}_{-0.009} $  & $ -0.96^{+0.03}_{-0.03} $    \\ 
\hline
\end{tabular}
\caption{Cosmology constraints from individual DSPL, combined DSPLs, DSPLs combined with complete CMB data, and DSPLs combined with CMB, SNe, and BAO data. Combined DSPL constraints use the updated constraints from J0946 obtained using spectroscopically confirmed redshift for the farther (i.e., second) source. Constraints for the flat \lcdm \ and flat \wcdm \ models are visualized in \autoref{fig:Om_LCDM} and \autoref{fig:wCDM}, respectively.
Final joint constraints after combining all four independent observations are shown \autoref{fig:wCDM all}.
See \autoref{Sec:Comological inferences} for a detailed discussion on the cosmological inference from the individual and combined observations. }
\label{tab:cosmology}
\end{table}

\subsection{Combined constraints with other standard probes}
\label{subsec: cosmology conbined with other probes}

\begin{figure}
    \includegraphics[clip=true, trim= 02mm 02mm 0mm 0mm, width=0.48\textwidth]{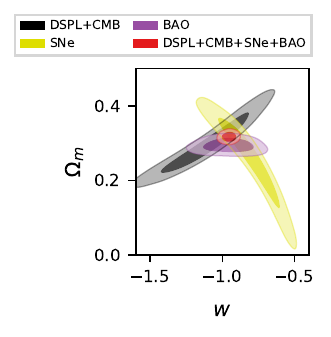}
    \caption{ 2D distribution of \wo\ and \om \ for the flat \wcdm\  model of the cosmology plotted using joint DSPL plus CMB observations (black contours), Type Ia SNe observations (yellow contours), BAO observations (purple contours), and all observations combined (red contours). See \autoref{Sec:Comological inferences} for more details.
    }
    \label{fig:wCDM all}
\end{figure}

Observations of CMB anisotropies, Type Ia SNe distances, and BAO constitute some of the most powerful probes of cosmology.
However, combining these datasets presents significant challenges to the standard \lcdm \ model, especially regarding the nature of dark energy  \citep{Perivolaropoulos:Skara:2022, DESI:DR2:BAO:2025}.
Constraints from a statistical sample of DSPLs are therefore valuable for providing independent and competitive tests of cosmological models, as well as for improving joint inference when combined with existing probes.
Using a mock sample of 87 DSPLs, \citet{Sharma:Collett:2023} demonstrated that DSPLs can constrain the dark-energy equation-of-state parameter with a precision comparable to that of the full \textit{Planck} CMB dataset.
While next-generation surveys--discussed in the following section--are expected to yield a substantial sample of DSPLs, this paper presents a pathfinder analysis based on only the second DSPL ever used for cosmography,   demonstrating their complementarity with existing cosmological probes.

Constraints from CMB observations alone, taken from the complete analysis by \citet{Planck:2020}, are plotted in the rightmost panel of \autoref{fig:wCDM} using pink contours.  
CMB observations alone infer \om$=0.315\pm 0.007$ for the flat \lcdm\ model and \om$=0.198^{+0.068}_{-0.038}$ and $w=-1.57^{+0.35}_{-0.27}$ for the flat \wcdm\ model.
The joint constraints from the two DSPLs, shown by the green contours in \autoref{fig:wCDM}, are perpendicular to the CMB constraints \citep[also observed with cluster lenses in][]{Caminha:Suyu:2022}. 
The orthogonality of the DSPL constraints to the CMB constraints is more clearly seen with a tighter contour from the larger mock sample of DSPLs in \citet{Sharma:Collett:2023}.

Following \autoref{eq:combined pdf}, the joint DSPL plus CMB constraints are found to be \om$=0.293^{+0.060}_{-0.050}$ and $w=-1.10^{+0.20}_{-0.22}$ for the flat \wcdm\ model.  
The final constrained region from joint DSPL and CMB observations is shown by the black contours in \autoref{fig:wCDM} and also in \autoref{fig:wCDM all}.
In \autoref{fig:wCDM all}, we also show the constraints from the Type Ia SNe dataset based on the DES-SN5YR observations \citep{DES:SNe:2024} and from BAO dataset based on DESI 1 year plus SDSS observations \citep{Adame:BAO:2025}, represented by the yellow and purple contours, respectively. 
Red contours in \autoref{fig:wCDM all}, represent the final constraints obtained by combining all four observations DSPLs, CMB, SNe, and BAO. The final combined constraints for the flat \wcdm\ model are \om$=0.316\pm0.009$ and \wo$=-0.96\pm0.03$.

Combining DSPL constraints with CMB observations shifts the posterior of \om\ and \wo\ by $\sim 1.2 \sigma$, bringing them closer to the concordance \lcdm\ model parameters, and improves the precision on \wo\ by $32\%$ compared to CMB data alone.  
The constraints from \agel1507 alone are broad, such that the $1\sigma$ credible region favors a large range (\om$ \lesssim 0.7 $ and \wo$\lesssim -0.5$).  
The constraints from J0946 alone overlap with the CMB constraints only at the $2\sigma$ level.  
However, the combined DSPL constraints (green curve) overlap with the CMB observations within the $1\sigma$ credible region and rule out the region favored by the CMB constraints where \om\ and \wo\ are small, demonstrating the complementarity of DSPLs with CMB for cosmography.  

Furthermore, combining the DSPL and CMB constraints with those from SNe and BAO observations constrains \om\ and \wo\ with a precision of approximately $3\%$ in the \wcdm\ model. 
We note that, given the use of only two DSPLs, 
most of this constraining power currently comes from the other probes. However, DSPL remains a highly promising probe that can independently deliver substantial constraining power on the cosmological parameters, thereby further improving joint inferences as the DSPL sample grows in the coming years \citep[see forecasts from][]{Sharma:Collett:2023, Shajib:Rubin-forecast:2024}.

\subsection{Future scope}
\label{sec:future scope}
With only a sample of two DSPLs, we limit this paper to testing the flat \lcdm \ and \wcdm \ models only. However, with a larger sample of DSPLs expected to be discovered by next-generation surveys such as \textit{Euclid} and the Rubin LSST, this work can be further expanded to constrain the curvature of the Universe (\ok) and the evolution of the dark-energy equation-of-state parameter with redshift \citep[e.g., with a redshift-dependent parametrization for $w$ as $w(z)=w_0 +w_a z/(1+z)$,][]{Chevallier:Polarski:2001, Linder:2003}.

Assuming a resolution of $0.12\arcsec$ and an $I$-band depth of 27 mag, \citet{Gavazzi:Treu:2008} predicted the detection of one DSPL per 40–80 galaxy-scale lenses. The spectroscopic  \agel\ survey has detected six DSPLs out of 100 confirmed lenses so far \citep{AGEL:DR2:2025}. \agel1507, studied here, is one of these six DSPLs, and the modeling of another DSPL, \agel035346-170639, is in progress (D. Bowden et al., in preparation). 
The addition of constraints from \agel035346-170639 is expected to improve the precision of the combined DSPL constraints by at least $\sim 5$–$10\%$, assuming it has similar sensitivity to cosmological parameters as \agel1507.

The ongoing \textit{Euclid} observations are expected to deliver a sample of 1200–1700 DSPLs suitable for cosmology \citep{Euclid:Li:2025} and $\sim 40$ rare triple-source-plane lenses \citep{Collett:Auger:2014}. 
In the first quick data release, \textit{Euclid} survey has identified 4 DSPLs out of 500 galaxy-scale lenses \citep{Euclid:Walmsley:2025}.
The upcoming Rubin LSST is expected to detect a sample of 500 DSPLs \citep{LSST:DES:2018, Shajib:Rubin-forecast:2024}. 
The follow-up 4MOST Strong Lens Spectroscopic Legacy Survey \citep[4SLSLS;][]{Collett:4MOST:2023} is expected to 
confirm a sample of $\sim 300$ DSPLs by providing 
spectra for source and deflector redshift measurements. 
This data will also be valuable to measure stellar kinematics to address the mass-sheet degeneracy. Such a sample will provide an $8\%$ precision on the inferred $w$ parameter using the DSPL sample alone \citep{Collett:4MOST:2023}. 

Assuming a $1\%$-level measurement of the $\beta$ parameter (similar to this work) for a mock sample 87 LSST DSPLs,  \citet{Sharma:Collett:2023} demonstrate DSPLs' potential in independently constraining dark energy parameters ($w_0, w_a$) and \ok. 
This sample represents a lower limit on expected detections in LSST’s best-seeing single-epoch imaging with follow-up spectroscopy \citep{LSST:DES:2018}.
Using the same DSPL sample alongside other cosmological probes in the LSST data, \citet{Shajib:Rubin-forecast:2024} refine the forecast for LSST constraints on dark energy, showing that strong lensing will be one of the most powerful dark energy probes from the Rubin LSST. 

Finally, even with a limited sample of just two DSPLs, this work demonstrates that DSPLs are promising cosmological probes, capable of providing valuable constraints independently and in combination with other probes, especially the CMB. 
With future larger samples, joint constraints from DSPLs are expected to achieve precision comparable to that of standard probes of cosmology.
A statistical sample of DSPLs will be a powerful tool for addressing current challenges to the \lcdm\ model, constraining the nature of dark energy, and guiding the development of more flexible models to accommodate observations across all scales.

\section{Conclusion}
\label{Sec:conclusions}
Inference of cosmological parameters from independent observations is important for testing the concordance \lcdm\ model of the Universe which faces many challenges observationally \citep{Perivolaropoulos:Skara:2022, DESI:DR2:BAO:2025}.
Galaxy-scale double-source-plane lenses, with two sources at widely separated redshifts, are a powerful probe of cosmology. Lens modeling of DSPLs provides an independent measurement of the cosmological distance ratio, $\beta_{12}$, which can be used to constrain cosmological parameters such as \om\ and $w$, independent of Hubble's constant (\autoref{Sec:compound lens theory}). In DSPLs, the high-$z$ lensed source offers additional constraints on the deflector mass profile, enabling robust lens modeling.

In this paper, we demonstrated the use of the DSPL \agel1507\ to independently constrain cosmological parameters in the flat \lcdm\ as well as more flexible flat \wcdm\ cosmological models.
We showed the improvement in constraints when combining those from \agel1507\ with another such lens, J0946, modeled by  \citet{Collett:Auger:2014}, effectively doubling the sample of DSPLs used for cosmological inference thus far. Importantly, we showed that constraints from DSPLs improve the final cosmological inference when combined with independent standard probes such as the CMB, Type Ia SNe, and BAO observations, highlighting the complementarity of DSPLs with other cosmological probes, especially CMB observations. 

We used the HST/WFC3 F140W-band image of \agel1507\ for lens modeling (\autoref{Sec:data}). We also used Keck/KCWI integral field spectroscopic observations to measure the redshifts of the two background sources. Additionally, we used the integrated spectrum of Source 1 to measure its stellar velocity dispersion, which was required to test our final lens model. 
The redshift and stellar velocity dispersion of the main deflector galaxy in \agel1507\ were already established by SDSS-BOSS.

We modeled the DSPL \agel1507\ using \textsc{lenstronomy} in multi-plane mode (\autoref{Sec:modeling}). In our lens model, we used a composite mass profile for the main deflector galaxy, comprising both dark matter and stellar mass components. Additionally, we accounted for the lensing effects of two small satellite galaxies, assuming they were in the deflector plane (\zdefl$=0.594$), and included Source 1 (\zone$=2.163$) as a deflector for Source 2 (\ztwo$=2.591$). Our final lens model, shown in \autoref{fig:model_composite_F140W}, successfully reconstructed the naked-cusp lensed configuration of Source 1, the quad lensed configuration of Source 2, and the intrinsic images of both sources.

The mass model of the main deflector galaxy predicted a velocity dispersion of $\sigma_{\rm pred}=290\pm 30$ km\,s$^{-1}$, consistent with the observed value of $\sigma_{\rm obs}=303\pm 38$ km\,s$^{-1}$ from the SDSS-BOSS survey within the $1\sigma$ bound. We found a non-zero Einstein radius of $0\farcs044^{+0.042}_{-0.029}$ for Source 1, equivalent to $\sigma_{\rm pred}=97\pm 33$ km\,s$^{-1}$, suggesting that Source 1 also contributed to the lensing of the more distant Source 2, forming a compound lens. 
Importantly, we found that $\sigma_{\rm pred}$ for Source 1 is consistent with our observed value of $\sigma_{\rm obs}=109\pm27$ km\,s$^{-1}$ within the $1\sigma$ uncertainty bound.

The consistency of our lens model predicted velocity dispersions for the main deflector, and Source 1, along with those from independent observations, suggests that our final lens model is robust. This indicates that our choice of mass profile for the deflectors is close to the true mass profile and the effect of the mass-sheet degeneracy is limited on our cosmological result. 
Our lens model constrains the cosmological scale factor in \agel1507\ to be $\beta_{12}=0.953^{+0.008}_{-0.010}$. Details of the modeling results are described in \autoref{Sec:modeling results} and summarized in \autoref{table:deflector properties} and \autoref{table:model parameters}.

From \agel1507\ alone, we inferred \om$=0.334^{+0.380}_{-0.234}$ for the flat \lcdm\ model, and \om$=0.344^{+0.384}_{-0.239}$ with $w=-1.24^{+0.66}_{-0.51}$ for the flat \wcdm\ model of cosmology. 
Combining the constraints from \agel1507\ with the updated constraints from J0946 improved the precision of \om\ in the \lcdm\ model and the precision of \wo\ in the \wcdm\ model by 10\%, compared to \agel1507\  alone. This improvement is 15\% and 30\% compared to J0946 alone.

We observed that the joint DSPL constraints are orthogonal to the CMB constraints.
Combining constraints from the two DSPLs with those from the CMB observations yielded \om$=0.293^{+0.060}_{-0.050}$ and $w=-1.10^{+0.20}_{-0.22}$ in the flat \wcdm\ model. 
Combined inferences, albeit with only two DSPLs, shifted the  median values of \om\ and \wo\ by approximately $1.2\sigma$, bringing them closer to the concordance \lcdm \ cosmology, and 
improved the precision of the \wo\ parameter by $30\%$ compared to CMB observations alone. 
This demonstrates that DSPLs are promising probes of cosmography and highly complementary to the other standard probes, especially CMB observations.
Individual and joint cosmological constraints are discussed in \autoref{Sec:Comological inferences} and summarized in \autoref{tab:cosmology}. 

With a total of two DSPLs, we limited the scope of this paper to constraining only the flat \lcdm\ and \wcdm\ models of cosmology. 
A larger sample of DSPLs is needed to test more flexible models by also constraining the evolution of dark energy equation-of-state parameters and the curvature of the Universe. 
In follow-up papers, we will increase the number of DSPLs used for cosmography by modeling confirmed DSPLs from the \agel \ survey. 
Based on current forecasts for \textit{Euclid} and the Rubin LSST observations (\autoref{sec:future scope}), along with the ongoing \agel\ survey, by the late 2020s, we are expected to have at least  $\sim 100$ DSPLs suitable for cosmography, providing inferences with precision comparable to that of other standard cosmological probes.

\begin{acknowledgements}
This research was supported by the Australian Research Council Centre of Excellence for All Sky Astrophysics in 3 Dimensions (ASTRO 3D), through project number CE170100013. 
NS thanks Giovanni Ferrami for helpful discussions on lens population forecasting.
%AJS
Support for this work was provided by NASA through the NASA Hubble Fellowship grant HST-HF2-51492 awarded to AJS by the Space Telescope Science Institute (STScI), which is operated by the Association of Universities for Research in Astronomy, Inc., for NASA, under contract NAS5-26555. AJS and HS were supported by NASA through the STScI grant HST-GO-16773, and AJS also received support through the STScI grant JWST-GO-2974.
%TJ
TJ, SR, and KVGC gratefully acknowledge financial support from NASA through grant HST-GO-16773, the Gordon and Betty Moore Foundation through Grant GBMF8549, the National Science Foundation through grant AST-2108515, and from a UC Davis Chancellor's Fellowship.
%KVGC
KVGC was supported by NASA through the STScI grants JWST-GO-04265 and JWST-GO-03777.
SHS thanks the Max Planck Society for support through the Max Planck Fellowship. 
T.N. acknowledges support from the Australian Research Council Laureate Fellowship FL180100060.

Other software used: \textsc{Numpy} \citep{NumPy:2020}, \textsc{Scipy} \citep{SciPy:2020}, \textsc{Astropy} \citep{Astropy:2013,  Astropy:2018, Astropy:2022}, \textsc{Matplotlib} \citep{matplotlib:2007}, 
 \textsc{ChainConsumer} \citep{Hinton2016}, \textsc{emcee} \citep{Foreman-Mackey:Hogg:2013}, 
 \textsc{GetDist} \citep{Lewis:2019},
\textsc{Multiprocess} \citep{Multiprocess:McKerns:Strand:2012}
\end{acknowledgements}

\bibliographystyle{aasjournal}
\bibliography{Lens_bibliography}

\appendix
\renewcommand{\thefigure}{A\arabic{figure}}
\renewcommand{\thetable}{A\arabic{table}}
\setcounter{figure}{0}
\setcounter{table}{0}

\section{Lens model parameters for \agel1507}
\label{appendix: model parameters}

\autoref{table:model parameters} summarizes the model parameters for each component in the lens model of DSPL \agel1507, obtained using \textsc{lenstronomy}. The table lists the median values from the final, converged MCMC chain, along with the $1\sigma$ ($68\%$) credible intervals spanning the 16th to 84th percentiles. \autoref{Sec:modeling} in the main paper describes the lens model components and the modeling process. Further details on the model profile parameterizations can be found in the \textsc{lenstronomy} \href{https://lenstronomy.readthedocs.io/}{documentation}.

\begin{table*}[h]
    \centering
    \begin{tabular}{|l|rrrrrrr|} \hline 
       \textbf{Components} &   \multicolumn{6}{c}{\textbf{Parameters}} & \\
        \hline
         \multicolumn{8}{l}{\textbf{Mass} }  \\
         \hline
         NFW Ellipse & \bm{$R_s$} (arcsec) & \bm{$\alpha_{Rs}$} & \bm{$e_{1}$} & \bm{$e_{2}$}& \bm{$x_0$}& \bm{$y_0$} & \\
         \cline{2-8}
          (main deflector, DG1)  &  $29.209^{+2.444}_{-2.293}$ &  $6.864^{+0.372}_{-0.315} $  & $ -0.359^{+0.006}_{-0.007}$  & $-0.047^{+0.003}_{-0.002}$ & $-0.800^{+0.001}_{-0.001}$ & $-0.481^{+0.001}_{-0.001} $ & \\ 
        \hline
        Double Chameleon & \bm{$M/L$} & \textbf{ amp ratio} & \bm{$w_{c1}$} (arcsec) & \bm{$w_{t1}$} (arcsec) & \bm{$e_{11}$}& \bm{$e_{21}$} & \\
         \cline{2-8}
           (DG1)  & $ 0.392^{+0.045}_{-0.053}$ & $0.976^{+0.010}_{-0.018}$  &  $0.08^{+0.001}_{-0.002}$ & $0.341^{+0.007}_{-0.006}$ & $-0.034^{+0.003}_{-0.002}$ & $-0.014^{+0.001}_{-0.001 }$  & \\ 
        \cline{2-8}
          & \bm{$w_{c2}$} (arcsec) & \bm{$w_{t2}$} (arcsec) & \bm{$e_{12}$} & \bm{$e_{22} $}& \bm{$x_0$}& \bm{$y_0$} & \\
         \cline{2-8}
            & $ 0.337^{+0.003}_{-0.008}$  & $ 2.966^{+0.006}_{-0.003}$ & $ -0.361^{+0.001}_{-0.001}$ & $ -0.017^{+-0.001}_{--0.001}$ & $-0.800^{+0.001}_{-0.001}$& $-0.481^{+0.001}_{-0.001} $ &  \\ 
           \hline
         Residual (or, external shear) & \bm{$\gamma_{\rm ext}$} & \bm{$\varphi_{\rm ext}$} & \multicolumn{4}{l}{ } &\\
           \cline{2-8}
             & $0.098^{+0.007}_{-0.007}$ & $1.470^{+0.015}_{-0.021}$ &  \multicolumn{4}{l}{ } &\\
           \hline 
         SIS & \bm{$\theta_{\rm E}$} (arcsec) &  \bm{$x_0$}& \bm{$y_0$} & \multicolumn{3}{l}{ } &\\
           \cline{2-8}
         (DG2)    & $0.130^{+0.019}_{-0.019}$ & $-0.411^{+0.002}_{-0.002}$ & $1.145^{+0.004 }_{-0.003}$ & \multicolumn{3}{l}{ } &\\
           \hline
        SIE & \bm{$\theta_{\rm E}$} (arcsec) & \bm{$e_{1}$} & \bm{$e_{2}$} & \bm{$x_0$}& \bm{$y_0$} & \multicolumn{1}{l}{ } &\\ 
           \cline{2-8}
           (DG3)  & $0.103^{+0.012}_{-0.012}$ & $0.490^{+0.007}_{-0.018}$ & $0.234^{+0.066}_{-0.065}$ & $-0.710^{+0.022}_{-0.021}$ & $-1.865^{+0.016}_{-0.012}$ & \multicolumn{1}{l}{ } &\\
           \hline
       SIE & \bm{$\theta_{\rm E}$} (arcsec) & \bm{$e_{1}$} & \bm{$e_{2}$} & \bm{$x_0$}& \bm{$y_0$} & \multicolumn{1}{l}{ } &\\
           \cline{2-8}
          (S1)   &$0.044^{+0.042}_{-0.029}$ & $0.026^{+0.013}_{-0.013}$ & $-0.316^{+0.014}_{-0.015}$ & $-1.073^{+0.007}_{-0.008}$ & $1.060^{+0.034 }_{-0.033}$& \multicolumn{1}{l}{ } &\\
           \hline
        \multicolumn{8}{l}{\textbf{Lens Light} }  \\
           \hline
        S\'ersic & \bm{$R_{\rm half}$} (arcsec) & \bm{$n$} & \bm{$x_0$} & \bm{$y_0$} & &  &  \\
           \cline{2-8}
          (DG2)   & $0.323^{+0.035}_{-0.028}$& $5.018^{+0.645}_{-0.510}$ & $-0.411^{+0.002}_{-0.002}$ & $1.145^{+0.004 }_{-0.003}$ & & &\\
           \hline
        S\'ersic ellipse & \bm{$R_{\rm half}$} (arcsec) & \bm{$n$} & \bm{$e_{1}$}  & \bm{$e_{2}$} & \bm{$x_0$} & \bm{$y_0$} &  \\
           \cline{2-8} 
          (DG3)  & $0.136^{+0.023}_{-0.024}$ &    $0.964^{+0.481}_{-0.319}$ &$0.490^{+0.007}_{-0.018}$ &  $0.234^{+0.066}_{-0.065}$ & $-0.710^{+0.022}_{-0.021}$ & $-1.865^{+0.016}_{-0.012}$ &\\
           \hline 
        \multicolumn{8}{l}{\textbf{Source Light} }  \\
           \hline
        S\'ersic ellipse & \bm{$R_{\rm half}$} (arcsec) & \bm{$n$} & \bm{$e_{1}$}  & \bm{$e_{2}$} & \bm{$x_0$} & \bm{$y_0$} &  \\
           \cline{2-8}
          (S1)  & $0.186^{+0.010}_{-0.010}$   
            & $2.341^{+0.294}_{-0.247}$ & $0.026^{+0.013}_{-0.013}$ & $-0.316^{+0.014}_{-0.015}$ & $-1.073^{+0.007}_{-0.008}$ & $1.060^{+0.034 }_{-0.033}$& \\
           \hline
        Shapelets & \bm{$n_{\rm Max}$} & \bm{$\beta$} & \bm{$x_0$} & \bm{$y_0$} &  &  &  \\
           \cline{2-8}
           (S1)  & $8$ & $0.026^{+0.001}_{-0.001}$  &    $-1.083^{+0.008}_{-0.007}$ & $1.011^{+0.033}_{-0.033}$ & & &\\
           \hline
        S\'ersic ellipse & \bm{$R_{\rm half}$} (arcsec) & \bm{$n$} & \bm{$e_{1}$}  & \bm{$e_{2}$} & \bm{$x_0$} & \bm{$y_0$} &  \\
           \cline{2-8}
           (S2)  &$0.057^{+0.002}_{-0.002}$ &  $0.538^{+0.087}_{-0.078}$ &    $0.049^{+0.029}_{-0.029}$ &    $0.316^{+0.030}_{-0.032}$ &    $-0.110^{+0.025}_{-0.022}$ &    $-0.318^{+0.034}_{-0.023}$ &\\
           \hline
    \end{tabular}
    \caption{Parameters for deflector mass distribution, external shear, deflector light profile, and source light profiles for the DSPL \agel1507. Centers are represented by $x_0$ and $y_0$.
    For the double Chameleon profile, `amp ratio' is the ratio of the amplitude of the first to the second single Chameleon profiles.}
    \label{table:model parameters}
\end{table*}

\end{document}